\documentclass[prapplied,twocolumn,superscriptaddress,floatfix]{revtex4-2}
\usepackage{amsmath}
\usepackage{amssymb}
\usepackage{graphicx}
\usepackage{color}
\usepackage[colorlinks=true,linkcolor=blue,citecolor=blue,urlcolor=blue]{hyperref}
\usepackage{placeins}
\usepackage{verbatim}
\newenvironment{figcontext}{\comment}{\endcomment}
\newcommand{\figref}[2]{Fig.~\hyperref[#1]{\ref*{#1}(#2)}}

\newif\ifarxiv
\arxivtrue  

\newcommand{\suppref}[1]{%
  \ifarxiv
    Supplemental Material, Sec.~\hyperref[#1]{\ref*{#1}}%
  \else
    Supplemental Material~\cite{SuppMat}%
  \fi
}

\makeatletter
\def\section{%
  \@startsection{section}{1}{\z@}%
    {0.5cm \@plus1ex \@minus .2ex}%
    {0.3cm}%
    {\normalfont\small\bfseries\centering}%
}%
\def\subsection{%
  \@startsection{subsection}{2}{\z@}%
    {0.5cm \@plus1ex \@minus .2ex}%
    {0.3cm}%
    {\normalfont\small\bfseries\centering}%
}%
\makeatother
\begin{document}

\title{Digital Predistortion of Optical Fields for Fast and High-Fidelity Entangling Gates in Trapped-Ion Qubits}

\author{Jovan~Markov$^{*,\dagger}$}
\affiliation{Department of Physics of Complex Systems, Weizmann Institute of Science, Rehovot 7610001, Israel}

\author{Yotam~Shapira$^{*}$}
\affiliation{Quantum Art, Ness Ziona 7403682, Israel}

\author{Ayelet~Hasson}
\affiliation{Department of Physics of Complex Systems, Weizmann Institute of Science, Rehovot 7610001, Israel}

\author{Meir~Alon}
\affiliation{Department of Physics of Complex Systems, Weizmann Institute of Science, Rehovot 7610001, Israel}

\author{Avraham~Gross}
\affiliation{Department of Physics of Complex Systems, Weizmann Institute of Science, Rehovot 7610001, Israel}

\author{Nitzan~Akerman}
\affiliation{Department of Physics of Complex Systems, Weizmann Institute of Science, Rehovot 7610001, Israel}

\author{Roee~Ozeri}
\affiliation{Department of Physics of Complex Systems, Weizmann Institute of Science, Rehovot 7610001, Israel}
\affiliation{Quantum Art, Ness Ziona 7403682, Israel}

\date{March 29, 2026}

\begin{abstract}
High-fidelity quantum gates require precise classical control signals, yet the analog hardware delivering these signals introduces nonlinear distortions that degrade gate performance. We demonstrate digital predistortion of an acousto-optic modulator used to generate multi-tone entangling-gate waveforms in a trapped-ion processor based on $^{88}$Sr$^+$. By measuring and inverting the static nonlinear amplitude response of the modulator, we apply a feed-forward correction that extends its linear operating range and suppresses spurious intermodulation products. Spectral analysis of the gate beam shows 3--5\,dB suppression of the dominant intermodulation tones, approximately doubling the usable diffraction efficiency at a $10^{-3}$ estimated gate-error threshold. Direct two-qubit Bell-state fidelity measurements confirm that predistortion consistently improves entangling-gate performance. The calibrate-and-invert methodology is device and platform agnostic, applicable to any nonlinear element in the classical control chain of a quantum processor.
\end{abstract}

\maketitle

\noindent $^{*}$These authors contributed equally to this work.\\
$^{\dagger}$Corresponding author: jovan.markov@weizmann.ac.il

\section{Introduction}

The performance of quantum gates depends not only on qubit coherence and control protocol design, but also on the faithfulness with which classical analog hardware delivers the intended waveforms. 
Across quantum computing platforms, nonlinear elements in the classical drive chain, including acousto-optic modulators (AOMs) in laser-driven trapped-ion gates~\cite{Leibfried2003,Gaebler2016}, radio-frequency (RF) power amplifiers in microwave~\cite{Ospelkaus2011,Zarantonello2019,Weber2024} and magnetic-gradient~\cite{Mintert2001,Srinivas2021,Nunnerich2025} gates, and electro-optic modulators~\cite{Neumuller2024}, generate spurious intermodulation products that distort the intended driving field and degrade gate performance~\cite{Manovitz2022,Gazalet1993,Saleh1981}.
Linear pulse distortions, arising from frequency-dependent attenuation and dispersion in the control chain, have been addressed through \textit{in situ} characterization and digital filtering techniques in superconducting~\cite{Gustavsson2013,Jerger2019,Rol2020,Guo2024,Hellings2025,Aggarwal2025} and spin-qubit~\cite{Ni2025,Duan2026} platforms, but nonlinear distortions remain largely uncompensated.
These nonlinearities are especially damaging for the spectrally dense multi-tone waveforms used in programmable entangling gates~\cite{Shapira2018,Hayes2012,Multiqubit2020,Sapiroid2023,Peleg2023,Schwerdt2024,Solomons2025}, where they produce intermodulation tones that can fall directly on motional sideband frequencies~\cite{Manovitz2022}, creating a practical trade-off between gate speed and fidelity.

Here we address this by adapting digital predistortion (DPD), a feed-forward linearization technique widely used in RF communications~\cite{Nagata1989,Cavers1990,Eun1997,Morgan2006,Ghannouchi2009,Cripps2006}, to the optical drive chain of a trapped-ion quantum processor.
By calibrating the static nonlinear amplitude response of the AOM and applying its numerical inverse, we extend the effective linear operating range, suppressing intermodulation products and enabling higher-power operation without fidelity loss.
We demonstrate this on multi-tone entangling-gate waveforms for $^{88}$Sr$^+$ ions and benchmark the improvement through both spectral analysis of the gate beam and direct two-qubit Bell-state fidelity measurements.
Although amplitude linearization of optical modulators has been demonstrated as a general instrumentation technique~\cite{Neumuller2024,Liu2023_OE}, phase predistortion has improved single-qubit trapped-ion gates~\cite{Gely2024}, and memory-polynomial DPD has been simulated for RF power amplifiers~\cite{Jaeger2026}, nonlinear predistortion has not previously been demonstrated to improve entangling-gate fidelity.
Because the method requires only a measured transfer function and its numerical inverse, it is directly transferable to other nonlinear elements in the drive chain of any qubit platform.

\section{Methods}

\subsection{Digital predistortion scheme}
\label{sec:dpd-scheme}

The goal of DPD is to linearize the effective transfer function of the various amplifiers and modulators used to drive the qubits. In this work we focus on the AOM that modulates the intensity of a first-order diffracted laser beam according to the desired multi-tone gate waveform. An AOM that is driven with high RF power to maximize diffraction efficiency presents a compressive nonlinear response at high drive amplitudes. In the absence of correction, the AOM maps the instantaneous RF drive envelope \(x(t)\) to the optical field envelope \(y(t)\) through the nonlinear transfer function. Here, we define \(y(t)\) as the optical field-amplitude envelope proportional to the square root of the optical power, \(|y|\propto\sqrt{P_{\mathrm{opt}}}\).
\begin{equation}
y(t)=f\!\big(x(t)\big) .
\label{eq:transfer-function}
\end{equation}
To compensate this distortion, we apply the inverse function to the desired waveform $x_{\mathrm{ideal}}(t)$,
\begin{equation}
x_{\mathrm{DPD}}(t)=f^{-1}\!\big(x_{\mathrm{ideal}}(t)\big),
\end{equation}
so that the optical output reproduces the target gate envelope, \(y(t)\approx x_{\mathrm{ideal}}(t)\).

\subsection{AOM characterization}
\label{sec:aom-characterization}

\begin{figure}[tbp]
\centering
\includegraphics[width=\columnwidth]{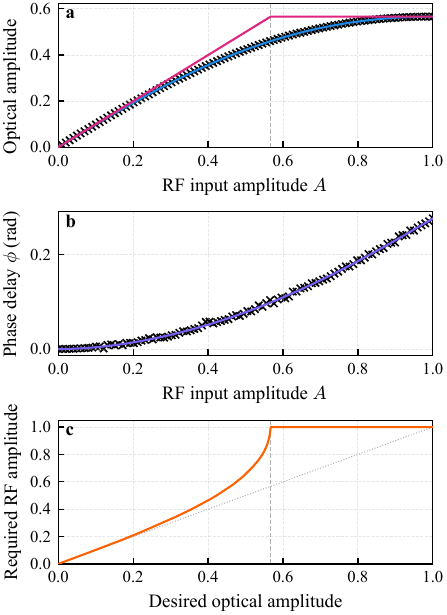}
\caption{AOM characterization and predistortion.
(a)~Amplitude transfer function: black crosses show photodiode-measured optical amplitude versus normalized RF input amplitude \(A\); blue line is the polynomial fit (Eq.~\eqref{eq:amplitude-poly}); magenta line shows the effective AOM output when DPD is applied, demonstrating linearity up to \(A_{\mathrm{corr}}\). Vertical dashed line: \(A_{\mathrm{corr}}\approx0.566\).
(b)~Phase transfer function: black crosses show the phase delay \(\phi(A)\) measured by heterodyne detection; purple line is the polynomial fit. The amplitude-dependent phase shift is not corrected by the amplitude-only DPD applied in this work.
(c)~Predistortion function: the orange curve gives the RF input amplitude required to produce a desired optical output amplitude, obtained by numerically inverting the polynomial fit in~(a). Gray dotted diagonal: the \(y=A\) line expected for a perfectly linear AOM, where no predistortion would be needed. Vertical dashed line: \(A_{\mathrm{corr}}\), at which the predistortion reaches full drive (\(A=1\)); for higher target amplitudes the correction has no additional headroom and the drive is clamped.}
\label{fig:aom_response}
\end{figure}

In an AOM, an applied RF drive creates an acoustic wave that diffracts part of the input laser beam into a first-order output whose intensity nominally follows the drive amplitude envelope.
In our setup, a small pick-off of this first-order diffracted beam was directed onto a fast photodiode whose voltage, integrated over a fixed time window, provided a signal proportional to the optical power of the output laser beam.
For each RF envelope amplitude \(x\), around a carrier at the AOM center frequency of \(\sim90~\mathrm{MHz}\), the corresponding photodiode signal was background-subtracted and integrated to obtain the corresponding optical power \(P_{\mathrm{opt}}\), from which we extract the optical amplitude, $y$, shown in \figref{fig:aom_response}{a}.

We introduce a normalized RF drive amplitude,
\[
A \equiv \frac{x}{x_{\mathrm{max}}},
\]
where \(x_{\mathrm{max}}\) is the RF amplitude at which the first-order diffraction efficiency reaches its maximum.
Thus \(A=1\) sets the upper end of the measured input range.
We further normalized the measured optical power such that the AOM response curve has unit slope at the origin; henceforth $y$ denotes this normalized optical amplitude. In these normalized coordinates, the transfer function of Eq.~\eqref{eq:transfer-function} becomes $y(A)$, and an ideal linear modulator satisfies $y = A$ for all $A\in[0,1]$.

The DPD correction maps each desired output amplitude to the RF input that produces it on the ideal linear curve. In practice, the DPD correction can be done only up to a maximum correctable amplitude \(A_{\mathrm{corr}}<1\) (vertical dashed line in \figref{fig:aom_response}{a}), at which the predistorted input reaches the full drive amplitude \(A=1\).
We find \(A_{\mathrm{corr}}\) as the point where the unit-slope line \(y=A\) reaches the maximum measured output \(y(1)\); since the response is normalized to unit slope at the origin, \(A_{\mathrm{corr}} = y(1)\), and the resulting window \(0\le A\lesssim A_{\mathrm{corr}}\) defines the correctable range.

We fit the measured amplitude with a polynomial model,
\begin{equation}
y(A)=\sum_{k=1}^{K}a_k A^k ,
\label{eq:amplitude-poly}
\end{equation}
where the \(a_k\) are real coefficients and \(K=8\) is the polynomial order.
The resulting amplitude transfer function is numerically inverted to obtain the predistortion mapping used in the experiments.

The phase response of the AOM, \(\phi(A)\), was measured by heterodyne detection of the optical beat between the diffracted gate beam and a single-frequency reference beam offset by \(\sim20~\mathrm{MHz}\), and is shown in \figref{fig:aom_response}{b}; calibration details are given in the \suppref{sec:supp-phase-calibration}.
Together these define a complex AOM transfer function $Y(A)=y(A)e^{i \phi(A)}$, which is used in numerical simulations to model the full AOM response. The DPD correction we applied in our experiments, however, compensated only for the amplitude nonlinearity $y(A)$ (\figref{fig:aom_response}{a}); the resulting predistortion function, obtained by numerically inverting this response, is shown in \figref{fig:aom_response}{c}.

\subsection{Gate protocol and benchmarking metrics}
\label{sec:gate-protocol}

We define the instantaneous diffraction efficiency \(\eta(t)\) as the fraction of input laser power transferred to the first-order diffracted beam, i.e.\ \(\eta = \eta_{\mathrm{ref}}\,y^2(A)/y^2(1)\), where \(y(A)\) is the normalized optical amplitude defined in Sec.~\ref{sec:aom-characterization} and \(\eta_{\mathrm{ref}}=0.80\) is the measured peak first-order diffraction efficiency at full drive (\(A=1\)).
This provides a physically meaningful axis that relates RF drive to usable optical power.
For time-varying gate waveforms, we use the time-averaged diffraction efficiency,
\[
\bar{\eta} = \frac{1}{T}\int_0^T \eta(t)\,dt,
\]
where $T$ is the gate duration.

We benchmark DPD via optical pulses that realize two-qubit entanglement gates, on qubits encoded in trapped $^{88}\text{Sr}^+$ ions. The resulting gate fidelity is used as a proxy for the efficacy of DPD. The entanglement sequence is based on the Cardioid(1,2) entangling gate~\cite{Shapira2018}, which employs four tones: two blue-detuned at \(\omega_{b,j}=\omega_c+\nu+j\xi_0\) and two red-detuned at \(\omega_{r,j}=\omega_c-\nu-j\xi_0\) for \(j=1,2\), where \(\omega_c\) is the qubit carrier frequency, \(\nu\) is the motional mode frequency of the ion-crystal used to mediate the qubit-qubit interaction, and \(\xi_0\) an additional detuning that sets the gate rate.
This gate provides an ideal probe for AOM nonlinearity because, here, cubic nonlinearity ($a_3\neq0$) generates third-order intermodulation (IM3) products at
\begin{equation}
2\omega_{b/r,1}-\omega_{b/r,2}=\omega_c\pm\nu ,
\end{equation}
which fall exactly on resonance with the motional sidebands.
These spurious tones drive unintended spin-motion dynamics and directly degrade gate fidelity.
Suppressing them via DPD therefore provides a stringent and experimentally accessible test of the correction.

In addition to measuring gate fidelity directly on the ions, we independently characterize the optical spectrum emerging from the AOM using heterodyne detection.
The diffracted gate beam is interfered with a reference beam frequency-shifted by \(\sim20~\mathrm{MHz}\), producing a beat signal on a fast photodiode; demodulating this beat directly recovers the gate waveform spectrum (see \suppref{sec:supp-heterodyne} for details).
To quantify DPD performance we define the gate-tone-to-IM3 power ratio
\begin{equation}
R_{10}=10\log_{10}\!\left(\frac{P_1}{P_0}\right),
\end{equation}
where \(P_n\) denotes the power at detuning \(n\xi_0\) from the motional sideband: \(P_1\) is the first gate tone (at detuning \(\xi_0\), equal in power to the second gate tone at \(2\xi_0\) by design), and \(P_0\) is the third-order intermodulation product that falls directly on the motional sideband and is the dominant source of gate-fidelity loss.
A larger \(R_{10}\) indicates stronger relative power in the gate tone compared to the spurious IM3 product.
Comparing \(R_{10}\) with and without DPD directly reveals the intermodulation suppression achieved by predistortion.

The gate fidelity can be expressed analytically in terms of the frequencies, amplitudes, and phases of the drive tones~\cite{Shapira2018}.
This formalism extends naturally to include spurious tones: by incorporating the measured $P_0$ into the calculation, we obtain an estimated fidelity \(\mathcal{F}_{\mathrm{est}}(\bar{\eta})\) that quantifies the expected gate performance as a function of diffraction efficiency (see the \suppref{sec:supp-fidelity-estimation}).

\subsection{Experimental setup}
\label{sec:experimental-setup}

Experiments were performed on a two-ion $^{88}\mathrm{Sr}^+$ crystal confined in a linear Paul trap~\cite{Manovitz2022}.
Qubits were encoded in the $5S_{1/2}$ ($\left|0\right\rangle$) and metastable $4D_{5/2}$ ($\left|1\right\rangle$) levels, addressed by a narrow-linewidth ($\simeq 10$~Hz) 674~nm laser on the $S_{-1/2} \to D_{+1/2}$ transition ($\pi$ time $\simeq 3.8$~$\mu$s).
The Cardioid(1,2) gate was driven on the axial stretch mode ($\nu_{\mathrm{stretch}} \approx 1.84$~MHz), chosen for its slower heating rate.
Sideband thermometry on the stretch mode yielded an initial mean phonon number $\bar{n} \approx 0.1$ after cooling and a heating rate below $0.01$~quanta/ms. The Lamb-Dicke parameter for this mode is $\eta_{\mathrm{LD}} \approx 0.026$ (not to be confused with the diffraction efficiency $\eta$ defined above).
Multi-tone gate waveforms were generated by an arbitrary waveform generator (AWG), which drives the AOM (center frequency $\sim 90$~MHz) to imprint the DPD-corrected spectrum onto the 674~nm beam.
Gate durations ranged from $T_{\mathrm{gate}} = 1/\xi_0 \approx 150$--$310~\mu$s across the RF amplitude settings used ($A = 0.4$--$0.8$).
Gate fidelity was extracted from population and parity-fringe measurements (see the \suppref{sec:supp-fidelity-extraction}); DPD and no-DPD conditions were interlaced within each scan.
Each measurement used 625 experimental repetitions, and multiple scans at each RF amplitude were aggregated.
State readout used state-selective fluorescence detection on the 422~nm transition; reported fidelities are not corrected for state-preparation-and-measurement (SPAM) errors.

\section{Results}

\subsection{Spectral benchmarking}

Within the correctable range \(A\lesssim A_{\mathrm{corr}}\), under DPD the AOM response is effectively linear, and we expect both intermodulation suppression and gate-fidelity improvements.

\begin{figure}[tbp]
\centering
\includegraphics[width=\columnwidth]{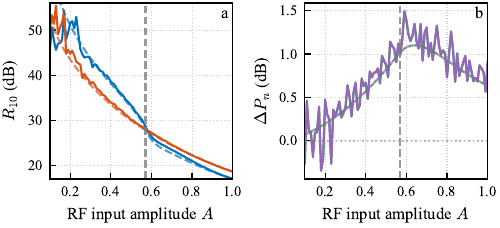}
\caption{Spectral benchmarking of amplitude-only DPD for the Cardioid(1,2) waveform.
Solid lines show heterodyne photodiode measurements; dashed lines show simulations incorporating the measured amplitude and phase transfer functions (Fig.~\ref{fig:aom_response}). In both panels, the vertical dashed line marks \(A_{\mathrm{corr}}\).
(a)~Gate-to-IM3 power ratio \(R_{10}\). Higher values indicate weaker IM3 tone relative to the gate tone. Within \(A\lesssim A_{\mathrm{corr}}\), DPD improves \(R_{10}\) by \(\sim\)3--5\,dB (blue) compared to uncorrected operation (red); for \(A\gtrsim A_{\mathrm{corr}}\) the DPD advantage diminishes as expected.
(b)~Gate-tone power change due to DPD, \(\Delta P_n = 10\log_{10}(P^{\mathrm{DPD}}_n/P^{\mathrm{NoDPD}}_n)\), versus \(A\), for the two gate tones \(n{=}1\) (at detuning \(\xi_0\), green) and \(n{=}2\) (at detuning \(2\xi_0\), purple). The two tones track each other closely, as expected from their equal design amplitudes.}

\label{fig:cardioid_spectrum_comparison}
\end{figure}

We first show the results of DPD on the photodiode measured spectrum of the Cardioid(1,2) gate.
\figref{fig:cardioid_spectrum_comparison}{a} shows the gate-tone-to-IM3 power ratio \(R_{10}\) versus normalized RF amplitude \(A\), for DPD (blue) and uncorrected operation (red). Within \(A\lesssim A_{\mathrm{corr}}\), DPD consistently raises \(R_{10}\) by \(\sim\)3--5\,dB, indicating that the IM3 products are suppressed relative to the gate tones. \figref{fig:cardioid_spectrum_comparison}{b} shows that this suppression is accompanied by a modest gain in gate-tone power.
Dashed lines in both panels show simulations that incorporate the measured amplitude and phase transfer functions (Fig.~\ref{fig:aom_response}), which closely reproduce the photodiode measurements (solid lines). For \(A\gtrsim A_{\mathrm{corr}}\), performance degrades as expected, and DPD no longer improves $R_{10}$.

\subsection{Gate fidelity versus gate rate}

\begin{figure}[tbp]
\centering
\includegraphics[width=\columnwidth]{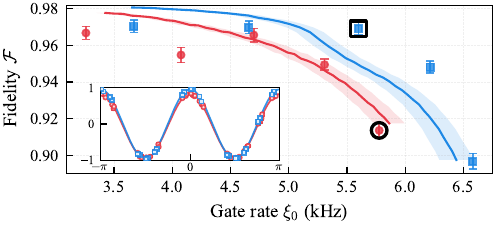}
\caption{Bell-state fidelity versus gate-rate parameter \(\xi_{0}\).
Solid lines: photodiode-derived fidelity estimates (aligned to gate data via two shared fit parameters; see text); shaded bands: gate-rate uncertainty from laser-power fluctuations.
Blue filled squares with error bars: measured gate fidelities with DPD; red filled circles with error bars: without DPD.
At a given gate rate, DPD consistently yields higher fidelity.
The highlighted data points (thick black outlines) mark two gates at similar rates (\(\xi_0 \approx 5.6\)\,kHz with DPD at $A=0.60$; \(\xi_0 \approx 5.8\)\,kHz without DPD at $A=0.80$), whose fidelities differ by \(\sim\)5.5 percentage points.
\textbf{Inset:} parity-fringe measurements for these two gates (blue open squares: DPD; red open circles: no~DPD; solid lines: sinusoidal fits). Fitted contrasts: $C_{\mathrm{DPD}} = 0.958 \pm 0.003$, $C_{\mathrm{NoDPD}} = 0.890 \pm 0.004$; the higher contrast with DPD shows that the fidelity gain is driven by improved coherence of the entangling operation.}
\label{fig:fidelity_vs_gate_rate}
\end{figure}

We next relate spectral purity to two-qubit gate performance.
Figure~\ref{fig:fidelity_vs_gate_rate} shows Bell-state fidelity versus the gate-rate parameter \(\xi_0\) for several drive amplitudes, with and without DPD.
The data points (markers) are fidelities measured directly from entangling-gate experiments, where \(\xi_0\) is precalibrated for each setting.
The solid curves are photodiode-derived fidelity estimates obtained by extracting the optical spectrum of the gate beam via heterodyne detection and computing the expected fidelity from the measured tone powers (see \suppref{sec:supp-fidelity-estimation}). To overlay these curves on the gate data, they are aligned using two shared parameters: a horizontal scale factor \(\alpha\) that converts relative gate-tone power to absolute \(\xi_0\), and a vertical fidelity offset \(\delta\) that accounts for systematic error sources independent of the AOM response (e.g., SPAM, dephasing, motional heating; see \suppref{sec:supp-two-param-fit}).
At a given gate rate, DPD consistently yields higher fidelity by suppressing the spurious intermodulation tones that otherwise degrade the gate.
The two highlighted data points illustrate this directly: at similar gate rates (\(\xi_0 \approx 5.6\)\,kHz with DPD at \(A=0.60\) and \(\xi_0 \approx 5.8\)\,kHz without DPD at \(A=0.80\)), DPD yields \(\sim\)5.5 percentage points higher fidelity, as confirmed by the parity-fringe contrast shown in the inset of Fig.~\ref{fig:fidelity_vs_gate_rate}.
The lower RF amplitude required with DPD is itself a consequence of the linearization: by suppressing intermodulation products, DPD channels more of the RF power into the intended gate tones, achieving the target gate rate at a less distorted operating point.
Agreement between the photodiode-derived fidelity estimates and the gate data holds for \(A\lesssim A_{\mathrm{corr}}\); above this threshold both degrade, consistent with the spectral breakdown seen in Fig.~\ref{fig:cardioid_spectrum_comparison}.

\begin{figcontext}
Context for "Fidelity versus gate rate (xi_0)" plot
================================================

What is shown
-------------
* Markers ("Gate"): experimental fidelities extracted from running the entangling gate and measuring the outcome.
  - For each gate point we calibrate xi_0 before running the gate.
  - xi_0 sets the gate time: T_gate = 1/xi_0 (i.e., we must know xi_0 to know how long to let the system evolve).

* Solid lines ("Photodiode"): fidelity estimates derived from heterodyne photodiode spectroscopy.
  - We interfere the multi-tone gate beam with a frequency-shifted single-tone reference on a fast photodiode.
  - Fourier analysis of the beat signal yields the optical spectrum of the gate (tone powers).
  - Using the extracted tone information, we compute the "appropriate fidelity" predicted for that measured spectrum, giving the fidelity curves (separately for DPD and no-DPD).

Why a gate-rate axis is nontrivial for the photodiode curves
-----------------------------------------------------------
* The gate data points naturally live on an absolute xi_0 axis because xi_0 is calibrated as part of running the gate.
* The photodiode-derived fidelities are obtained from spectra taken versus RF input settings (and DPD on/off), but do not directly provide an absolute xi_0 in kHz.
* Therefore, we need a mapping from "RF input amplitude + DPD flag" -> "gate rate xi_0".

Simulation-based mapping from RF amplitude to (relative) gate rate
------------------------------------------------------------------
We build a simulation that uses an AOM response model based on measured AOM calibrations:
* The model includes the AOM amplitude response calibration and the AOM phase response calibration.

For each RF input amplitude (scanned broadly from small values up to 1, and practically used in the 0.4-0.8 range where we took gate data) and for each DPD flag (0/1):
1) Synthesize the intended RF waveform (multi-tone gate waveform).
2) Optionally apply digital predistortion (DPD) to the waveform.
3) Pass the waveform through the calibrated AOM model to obtain the output waveform (optical-domain field in arbitrary units).
4) Take an FFT of the output to obtain its spectrum and extract the power in the intended gate tone.

Define an anchor and compute relative gate rate:
* Anchor choice: RF template amplitude = 0.4 with DPD = 0 (no DPD), because this is the lowest-rate gate point we measured.
* Gate-rate scaling assumption:
  - Gate rate is proportional to Rabi frequency.
  - Rabi frequency is proportional to sqrt(optical power in the relevant gate tone).
* Therefore, for each simulated setting we define a relative gate-rate factor:
      r_rel = sqrt( P_gate_tone / P_gate_tone(anchor) )
  so the anchor has r_rel = 1, and higher-power settings give r_rel > 1.

Putting photodiode fidelities on an absolute xi_0 axis
----------------------------------------------------
* The photodiode spectroscopy gives fidelity versus (RF amplitude, DPD flag).
* The simulation gives r_rel versus (RF amplitude, DPD flag).
-> Combining them produces a curve: Fidelity_photodiode versus r_rel.

To convert r_rel to an absolute gate rate in kHz, we introduce a single multiplicative scale factor:
      xi_0 = (scale in kHz) * r_rel
This leaves one free "horizontal" parameter (the kHz scale factor).

Fitting photodiode curves to gate data (two free parameters)
------------------------------------------------------------
To overlay the solid curves on the measured gate points, we fit using:
1) A horizontal scale factor that converts relative gate rate to absolute xi_0 in kHz.
2) A vertical fidelity offset (delta) between the photodiode-based estimator and the measured gate fidelities.
The fitted offset is delta = -0.018 +/- 0.003 (see Supplemental Material, Sec. S7).

Spread / band around the photodiode curves (gate-rate uncertainty from power fluctuations)
-----------------------------------------------------------------------------------------
* The solid line is based on the simulated mapping and/or a single nominal spectrum.
* In reality we have laser power fluctuations (e.g., from coupling fluctuations).
* We quantify the power fluctuation versus RF template by overlaying measured power-vs-template behavior with the simulation curve and extracting the spread relative to the simulation.

Key point: this is treated as a relative power fluctuation (power-independent in the sense that it is a fraction, e.g. "~10
      xi_0 is proportional to sqrt(P) => delta_xi_0/xi_0 = (1/2)*(delta_P/P)

As xi_0 increases (higher power), the absolute spread in xi_0 grows, which is why the band widens toward larger gate rates.
\end{figcontext}

\subsection{Gate fidelity versus diffraction efficiency}

\begin{figure}[tbp]
\centering
\includegraphics[width=\columnwidth]{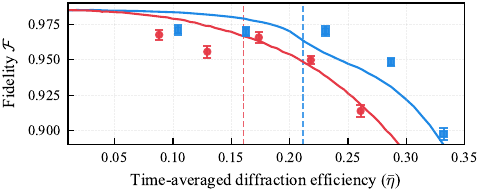}
\caption{Gate fidelity versus time-averaged diffraction efficiency \(\bar{\eta}\).
Blue solid line: photodiode-derived fidelity estimate with DPD; red solid line: without DPD.
Blue squares with error bars: measured gate fidelities with DPD; red circles: without DPD.
Vertical dashed lines mark \(\bar{\eta}_{\mathrm{corr}}\), the time-averaged diffraction efficiency obtained by driving the Cardioid(1,2) waveform at the maximum correctable amplitude \(A_{\mathrm{corr}}\), without DPD (red) and with DPD (blue).
Because DPD reshapes the RF envelope, the same input amplitude \(A_{\mathrm{corr}}\) yields a higher \(\bar{\eta}\) after predistortion, shifting the correctable operating range to higher optical throughput.}

\label{fig:fidelity_vs_eff}
\end{figure}

\begin{figure}[tbp]
\centering
\includegraphics[width=\columnwidth]{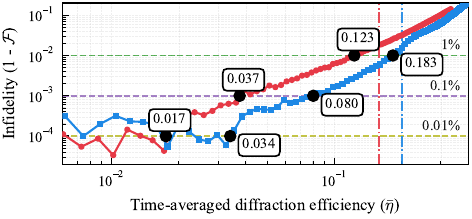}
\caption{Photodiode-based estimate of entangling-gate infidelity
$\epsilon_{\mathrm{est}} \equiv 1 - \mathcal{F}_{\mathrm{est}}$ versus time-averaged diffraction efficiency
$\bar{\eta}$ (log--log scale). $\mathcal{F}_{\mathrm{est}}$ is obtained by extracting the gate-tone and dominant IM3
powers from the heterodyne photodiode spectrum and evaluating the analytic gate-fidelity expression
(see \suppref{sec:supp-fidelity-estimation}).
Red solid line: without DPD; blue solid line: with DPD.
Horizontal dashed lines mark fixed error budgets:
$\epsilon_{\mathrm{est}}=10^{-2}$ (1\%, green), $10^{-3}$ (0.1\%, purple), $10^{-4}$ (0.01\%, yellow);
black dots with boxed labels give the corresponding threshold efficiencies $\bar{\eta}^{\mathrm{th}}$.
Vertical dash-dot lines mark $\bar{\eta}_{\mathrm{corr}}$, the time-averaged diffraction efficiency at the maximum correctable amplitude $A_{\mathrm{corr}}$, without DPD (red) and with DPD (blue).
DPD shifts every $\bar{\eta}^{\mathrm{th}}$ to higher $\bar{\eta}$; beyond the blue vertical line the DPD correction runs out of headroom and the advantage diminishes.}
\label{fig:pd_fidelity_vs_eff}
\end{figure}

To remove dependence on gate-rate calibration, we re-express performance in terms of the time-averaged diffraction efficiency \(\bar{\eta}\).
Figure~\ref{fig:fidelity_vs_eff} shows measured fidelities versus \(\bar{\eta}\) together with photodiode-based fidelity estimates.
DPD shifts the fidelity roll-off to higher \(\bar{\eta}\), allowing equivalent fidelity at higher optical power; the vertical dashed lines in Fig.~\ref{fig:fidelity_vs_eff} mark \(\bar{\eta}_{\mathrm{corr}}\) for each case, confirming that DPD extends the correctable range to higher throughput.
To resolve the low-infidelity regime and define quantitative operating thresholds, we replot the spectral estimate on a logarithmic scale.
A spectral benchmark (Fig.~\ref{fig:pd_fidelity_vs_eff}) converts the measured tone powers into an estimated infidelity \(\epsilon_{\mathrm{est}}(\bar{\eta}) \equiv 1 - \mathcal{F}_{\mathrm{est}}(\bar{\eta})\) and defines threshold efficiencies \(\bar{\eta}^{\mathrm{th}}(\epsilon)\) at fixed error budgets \(\epsilon_{\mathrm{est}}=10^{-2},10^{-3},10^{-4}\).
DPD increases \(\bar{\eta}^{\mathrm{th}}\) at every error budget: at \(\epsilon_{\mathrm{est}}=10^{-2}\), \(\bar{\eta}^{\mathrm{th}}\) rises from 0.12 without DPD to 0.18 with DPD (a factor of \(\approx 1.5\)), at \(\epsilon_{\mathrm{est}}=10^{-3}\) from 0.037 to 0.080 (a factor of \(\approx 2.1\)), and at \(\epsilon_{\mathrm{est}}=10^{-4}\) from 0.017 to 0.034 (a factor of \(\approx 1.9\)); at these lowest diffraction efficiencies the heterodyne signal-to-noise ratio is reduced and fidelity estimates become noisier, but the DPD improvement remains consistent with higher thresholds.
The vertical dash-dot lines in Fig.~\ref{fig:pd_fidelity_vs_eff} confirm that this gain persists up to \(\bar{\eta}_{\mathrm{corr}}\).

\begin{figcontext}
## Context for plot: Fidelity vs. time-averaged diffraction efficiency (bar_eta)

### What is shown
- **y-axis:** two-qubit gate fidelity F.
- **x-axis:** time-averaged diffraction efficiency bar_eta.

### Gate data (points with error bars)
- The data points come from actual gate experiments: we execute the entangling gate, measure the relevant populations (including parity fringes, etc.), and extract a fidelity for each setting.
- **Blue points:** gate data with DPD.
- **Red points:** gate data without DPD.

### Photodiode-derived prediction (solid curves)
- The solid curves are computed from an independent heterodyne photodiode diagnostic of the gate-beam spectrum.
- The spectral data are independent, but the curves shown here include the fitted vertical fidelity offset delta = -0.018 from the two-parameter fit (shared with Fig. 3; see Supplemental Material, Sec. S5.4). This offset accounts for error sources not captured by the spectral model (SPAM, dephasing, etc.).
- **Measurement:** we interfere (i) a single-frequency reference beam with (ii) the gate beam that contains the multi-tone gate spectrum, on a fast photodiode.
- **Analysis:** we take the photodiode beat signal and perform an FFT to obtain the optical spectrum.
- From the FFT spectrum we extract the powers in:
  1. the intended gate tones, and
  2. the most prominent third-order intermodulation products (IM3) generated by AOM nonlinearity.
- These extracted spectral components are inserted into our gate-fidelity equations/model (i.e., plug in the spectrum -> compute the expected fidelity), producing the curves:
  - **Blue curve:** spectrum-derived fidelity with DPD.
  - **Red curve:** spectrum-derived fidelity without DPD.

### Vertical dashed lines (bar_eta_corr)
- Two vertical dashed lines mark bar_eta_corr: the time-averaged diffraction efficiency at the maximum correctable amplitude A_corr.
- **Red dashed line:** bar_eta_corr without DPD.
- **Blue dashed line:** bar_eta_corr with DPD.
- Because DPD reshapes the RF envelope, the same A_corr yields a higher bar_eta after predistortion, so the blue line is to the right of the red line.

### Definition of bar_eta (x-axis)
- bar_eta is computed from the output waveform after the AOM.
- We obtain that output waveform numerically by propagating the programmed input waveform through a calibrated AOM response model (including the amplitude response and phase response calibrated beforehand).
- bar_eta is then calculated from the resulting output waveform and serves as a single-number summary of the average optical power throughput during the gate time.
- The reference diffraction efficiency is eta_ref = 0.80 (measured peak first-order efficiency at full drive).
\end{figcontext}


\begin{figcontext}
## Figure context: Photodiode-spectra-derived infidelity vs time-averaged diffraction efficiency

### What is plotted
* **Y-axis:** Estimated gate infidelity eps_est = 1 - F_est.
* **X-axis:** Time-averaged diffraction efficiency bar_eta (log-log).
  bar_eta is used as a proxy for the power throughput through the AOM (and therefore is closely related to the achievable gate rate / gate speed).
* The plot focuses on the "foot" of the data (low-infidelity region).
* **Blue:** AM-only DPD. **Red:** No DPD.

---

### How the infidelity curves are obtained (heterodyne photodiode diagnostic)
1. **Heterodyne interference on a fast photodiode**
   - Two optical beams are interfered: a single-frequency reference beam (frequency-shifted relative to the gate beam) and the gate beam, which contains the multi-tone spectrum (e.g., the Cardioid(1,2) spectrum).
2. **Beat-note acquisition**
   - The photodiode measures the beat signal, which lies within the detector bandwidth.
3. **Spectrum extraction**
   - An FFT of the measured beat signal yields the optical spectrum of the gate beam (via the heterodyne mapping).
4. **Extract relevant spectral components**
   - From the measured spectrum, extract: the powers in the intended gate tones and the dominant third-order intermodulation (IM3) products (the strongest distortion tones observed).
5. **Compute spectrum-derived estimated fidelity**
   - The extracted tone powers are used as inputs to an analytic multi-tone gate model to compute an estimated fidelity F_est.
   - This is an estimate of the gate fidelity from the measured spectrum alone, i.e., assuming no other error sources beyond what is captured by the spectral distortions.
6. **Convert to infidelity**
   - Plot eps_est = 1 - F_est.

---

### How bar_eta is defined / computed (time-averaged diffraction efficiency)
* bar_eta represents the AOM power throughput and is computed numerically.
* It is determined by: the RF input amplitude, and a DPD flag (off/on).
* Procedure:
  1. Generate the RF waveform numerically (for a given RF amplitude and DPD on/off).
  2. Pass it through an AOM model that includes the calibrated amplitude and phase response.
  3. From the modeled AOM output, compute the throughput, reported as bar_eta.

---

### Error-budget thresholds and the meaning of the boxed labels
* Horizontal dashed lines indicate example infidelity (error-budget) thresholds:
  - eps_est = 10^-2 (1
  - eps_est = 10^-3 (0.1
  - eps_est = 10^-4 (0.01
* For each threshold, the intersection of the dashed line with a curve defines the corresponding threshold efficiency bar_eta^th.
* The boxed numeric labels on the curves report these bar_eta^th values (the bar_eta where a given curve reaches the specified infidelity threshold).

---

### Interpretation: what improvement DPD provides
Two equivalent viewpoints:

**(1) Fixed error budget -> larger bar_eta (more throughput / faster gate)**
* At a fixed threshold (10^-2, 10^-3, 10^-4), the DPD curve reaches the threshold at a higher bar_eta than the No-DPD curve.
* This means that for the same allowed infidelity, DPD allows more AOM throughput (interpretable as enabling a faster gate / higher gate rate).

**(2) Fixed bar_eta -> lower error (higher fidelity)**
* At the same bar_eta (roughly the same throughput / gate-rate setting), DPD yields lower infidelity in the regime where it is effective.

---

### Low-bar_eta regime: noisy behavior and expected physical trend
* Around bar_eta ~ 0.02-0.03 and below, the curves become noisy.
* This is attributed to measurement limitations: at very low power, the noise floor reduces the SNR of the heterodyne/FFT spectrum extraction.
* Physically, at sufficiently low bar_eta the AOM is driven in a more linear regime, so there is little to predistort; therefore, one expects the DPD and No-DPD curves to merge in the very-low-bar_eta limit (even if the measured traces are noise-dominated there).

---

### High-bar_eta regime: diminishing DPD advantage
* At very high bar_eta, the DPD and No-DPD curves become similar again.
* This is consistent with DPD being effective only up to a point: correction can require sending in more power, and beyond some input/drive ceiling the distortions cannot be further compensated, so the improvement diminishes and the curves converge.
\end{figcontext}

\section{Conclusion}

We have demonstrated that amplitude-only digital predistortion (DPD) can linearize the static response of an acousto-optic modulator used to generate multi-tone entangling-gate waveforms in a trapped-ion processor.
By calibrating the AOM amplitude transfer function and applying its numerical inverse as a feed-forward correction, we extend the effectively linear operating range and suppress nonlinear intermodulation distortions.
Applying DPD to the Cardioid(1,2) entangling gate on $^{88}$Sr$^+$, we observe a 3--5\,dB suppression of the dominant spurious intermodulation tone relative to the gate tones, as measured by heterodyne spectroscopy of the laser beam delivering the gate waveform to the ions.
This spectral improvement translates into higher gate fidelity, as confirmed by parity-fringe measurements of two-qubit Bell states (Fig.~\ref{fig:fidelity_vs_gate_rate}, inset).
Furthermore, DPD shifts the onset of nonlinearity-limited performance to higher diffraction efficiencies, extending the range of optical throughput over which high-fidelity gates can be achieved.

The present implementation compensates only amplitude nonlinearities.
Future extensions include incorporating phase-response correction (\figref{fig:aom_response}{b}) to enable complex-valued (amplitude and phase) predistortion~\cite{Cavers1990, Guan2014}, and modeling history-dependent distortions using memory-polynomial DPD approaches~\cite{Kim2001, Ding2004, Morgan2006}, which may further improve performance at high modulation bandwidth.

Although demonstrated here on an AOM in a laser-driven gate, the calibrate-and-invert methodology is device and platform agnostic. The same procedure can linearize any nonlinear element in the classical drive chain of a quantum processor, including the RF power amplifiers used in microwave-driven trapped-ion~\cite{Ospelkaus2011,Zarantonello2019,Weber2024} and magnetic-gradient~\cite{Mintert2001,Srinivas2021,Nunnerich2025} gates, as well as electro-optic modulators~\cite{Neumuller2024} and other active devices in photonic and electronic control systems.
While operating at reduced drive power avoids nonlinear distortions, it does so at the cost of lower gate speed and optical throughput.
As quantum computing architectures scale to more qubits and increasingly complex gate waveforms, employing denser multi-tone spectra~\cite{Shapira2018,Multiqubit2020,Peleg2023,Schwerdt2024,Solomons2025} and higher drive powers, the intermodulation products generated by hardware nonlinearities will grow in number and strength, making feed-forward nonlinear predistortion an increasingly important tool for maintaining high-fidelity quantum control.

See the Supplemental Material for calibration details (Secs.~\hyperref[sec:supp-phase-calibration]{\ref*{sec:supp-phase-calibration}}--\hyperref[sec:supp-stability]{\ref*{sec:supp-stability}}), fidelity estimation (Sec.~\hyperref[sec:supp-fidelity-estimation]{\ref*{sec:supp-fidelity-estimation}}), and simulation methodology (Sec.~\hyperref[sec:supp-simulation]{\ref*{sec:supp-simulation}}).

\section*{Acknowledgements}
We thank Benjamin Nyagaka and Peter Szabo for helpful discussions. This work was supported by ISF Quantum grants 3475/21 and 1364/24, and by TWIN-Q Quantum computing center.

\newpage
\bibliography{bib/references}

@article{Sackett2000,
  author  = {Sackett, C. A. and Kielpinski, D. and King, B. E. and Langer, C. and Meyer, V. and Myatt, C. J. and Rowe, M. and Turchette, Q. A. and Itano, W. M. and Wineland, D. J. and Monroe, C.},
  title   = {Experimental entanglement of four particles},
  journal = {Nature},
  volume  = {404},
  pages   = {256--259},
  year    = {2000},
  doi     = {10.1038/35005011}
}

@mastersthesis{Manovitz2017MSc,
  title={Imaging and Individual Addressing Module for Trapped Ion Quantum Information Processing},
  author={Tom Manovitz},
  school={Weizmann Institute of Science},
  year={2017},
  url={https://www.weizmann.ac.il/complex/ozeri/sites/complex.ozeri/files/uploads/thesis_submission.pdf},
}

@misc{Neumuller2024,
  title={Bias Control and Linearization of the Transfer Function of Electro-optic and Acousto-optic Modulators},
  author={Neum\"{u}ller, Clemens and Obernosterer, Frank and Meyer, Raimund and Koch, Robert and Kilian, Gerd},
  year={2024},
  eprint={2403.10331},
  archivePrefix={arXiv},
  primaryClass={physics.optics}
}

@article{Shapira2018,
  title={Robust entanglement gates for trapped-ion qubits},
  author={Y. Shapira and R. Shaniv and T. Manovitz and N. Akerman and R. Ozeri},
  journal={Physical Review Letters},
  volume={121},
  number={18},
  pages={180502},
  year={2018},
  publisher={APS},
  doi={10.1103/PhysRevLett.121.180502}
}

@article{Manovitz2022,
  title={Trapped-Ion Quantum Computer with Robust Entangling Gates and Quantum Coherent Feedback},
  author={Tom Manovitz and Yotam Shapira and Lior Gazit and Nitzan Akerman and Roee Ozeri},
  journal={PRX Quantum},
  volume={3},
  pages={010347},
  year={2022},
  publisher={APS},
  doi={10.1103/PRXQuantum.3.010347}
}

@article{Multiqubit2020,
  title={Theory of robust multiqubit nonadiabatic gates for trapped ions},
  author={Y. Shapira and R. Shaniv and T. Manovitz and N. Akerman and L. Peleg and L. Gazit and R. Ozeri and A. Stern},
  journal={Physical Review A},
  volume={101},
  number={3},
  pages={032330},
  year={2020},
  publisher={APS},
  doi={10.1103/PhysRevA.101.032330}
}

@misc{Solomons2025,
  title={Full programmable quantum computing with trapped-ions using semi-global fields},
  author={Solomons, Yakov and Kadish, Yotam and Peleg, Lee and Nemirovsky, Jonathan and {Ben Kish}, Amit and Shapira, Yotam},
  year={2025},
  eprint={2509.14331},
  archivePrefix={arXiv},
  primaryClass={quant-ph}
}

@article{Schwerdt2024,
  title={Scalable Architecture for Trapped-Ion Quantum Computing Using rf Traps and Dynamic Optical Potentials},
  author={David Schwerdt and Lee Peleg and Yotam Shapira and Nadav Priel and Yanay Florshaim and Avram Gross and Ayelet Zalic and Gadi Afek and Nitzan Akerman and others},
  journal={Physical Review X},
  volume={14},
  pages={041017},
  year={2024},
  publisher={APS},
  doi={10.1103/PhysRevX.14.041017}
}

@misc{Peleg2023,
    title={Fast design and scaling of multi-qubit gates in large-scale trapped-ion quantum computers},
    author={Lee Peleg and David Schwerdt and Jonathan Nemirovsky and Yotam Shapira and Nitzan Akerman and Ady Stern and Amit Ben Kish and Roee Ozeri},
    year={2023},
    eprint={2307.09566},
    archivePrefix={arXiv},
    primaryClass={quant-ph},
    url={https://arxiv.org/abs/2307.09566},
}

@book{Goutzoulis1994,
  title={Design and Fabrication of Acousto-Optic Devices},
  editor={Goutzoulis, A. P. and Pape, D. R.},
  publisher={Marcel Dekker},
  address={New York},
  year={1994},
  doi={10.1201/9781003210221}
}

@article{Vernaleken2007,
  author={Vernaleken, A. and Cohen, M. G. and Metcalf, H.},
  title={Interferometric measurement of acoustic velocity in {PbMoO}$_4$ and {TeO}$_2$},
  journal={Applied Optics},
  volume={46},
  number={29},
  pages={7117--7119},
  year={2007},
  doi={10.1364/AO.46.007117}
}

@article{EsquivelRamirez2024,
  author={Esquivel-Ram\'{i}rez, E. and Vega-Hern\'{a}ndez, M. and Estudillo-Ayala, J. M. and Pottiez, O. and Jauregui-V\'{a}zquez, D. and Sierra-Hernandez, J. M. and Rojas-Laguna, R.},
  title={High-precision frequency-controlled optical phase shifter with acousto optic devices},
  journal={Optics Letters},
  volume={49},
  number={9},
  pages={2525--2528},
  year={2024},
  doi={10.1364/OL.522688}
}

@article{Gazalet1993,
  title={Acousto-optic multifrequency modulators: reduction of the phase-grating intermodulation products},
  author={Gazalet, M. G. and Kastelik, J. C. and Bruneel, C. and Bazzi, O. and Bridoux, E.},
  journal={Applied Optics},
  volume={32},
  number={13},
  pages={2455--2460},
  year={1993},
  doi={10.1364/AO.32.002455}
}

@article{Liu2023_OE,
  title={Using an acousto-optic modulator as a fast spatial light modulator},
  author={Liu, X. and Braverman, B. and Boyd, R. W.},
  journal={Optics Express},
  volume={31},
  pages={1501--1515},
  year={2023},
  doi={10.1364/OE.471910}
}

@article{Ghannouchi2009,
  title={Behavioral modeling and predistortion},
  author={Ghannouchi, F. M. and Hammi, O.},
  journal={IEEE Microwave Magazine},
  volume={10},
  number={7},
  pages={52--64},
  year={2009},
  doi={10.1109/MMM.2009.934516}
}

@book{Cripps2006,
  title={{RF} Power Amplifiers for Wireless Communications},
  author={Cripps, S. C.},
  edition={2nd},
  publisher={Artech House},
  address={Norwood, MA},
  year={2006}
}

@inproceedings{Nagata1989,
  title={Linear amplification technique for digital mobile communications},
  author={Nagata, Y.},
  booktitle={Proc. IEEE Vehicular Technology Conference (VTC)},
  pages={159--164},
  year={1989},
  doi={10.1109/VETEC.1989.40066}
}

@article{Saleh1981,
  author  = {Saleh, A. A. M.},
  title   = {Frequency-Independent and Frequency-Dependent Nonlinear Models of {TWT} Amplifiers},
  journal = {IEEE Transactions on Communications},
  volume  = {29},
  number  = {11},
  pages   = {1715--1720},
  year    = {1981},
  doi     = {10.1109/TCOM.1981.1094911}
}

@article{Cavers1990,
  title={Amplifier linearization using a digital predistorter with fast adaptation and low memory requirements},
  author={Cavers, J. K.},
  journal={IEEE Transactions on Vehicular Technology},
  volume={39},
  number={4},
  pages={374--382},
  year={1990},
  doi={10.1109/25.61359}
}

@article{Eun1997,
  author  = {Eun, C. and Powers, E. J.},
  title   = {A New {Volterra} Predistorter Based on the Indirect Learning Architecture},
  journal = {IEEE Transactions on Signal Processing},
  volume  = {45},
  number  = {1},
  pages   = {223--227},
  year    = {1997},
  doi     = {10.1109/78.552219}
}

@article{Kim2001,
  author  = {Kim, J. and Konstantinou, K.},
  title   = {Digital predistortion of wideband signals based on power amplifier model with memory},
  journal = {Electronics Letters},
  volume  = {37},
  number  = {23},
  pages   = {1417--1418},
  year    = {2001},
  doi     = {10.1049/el:20010940}
}

@article{Ding2004,
  author  = {Ding, L. and Zhou, G. T. and Morgan, D. R. and Ma, Z. and Kenney, J. S. and Kim, J. and Giardina, C. R.},
  title   = {A robust digital baseband predistorter constructed using memory polynomials},
  journal = {IEEE Transactions on Communications},
  volume  = {52},
  number  = {1},
  pages   = {159--165},
  year    = {2004},
  month   = jan,
  doi     = {10.1109/TCOMM.2003.822188}
}

@article{Morgan2006,
  title={A generalized memory polynomial model for digital predistortion of {RF} power amplifiers},
  author={Morgan, D. R. and Ma, Z. and Kim, J. and Zierdt, M. G. and Pastalan, J.},
  journal={IEEE Transactions on Signal Processing},
  volume={54},
  number={10},
  pages={3852--3860},
  year={2006},
  doi={10.1109/TSP.2006.879264}
}

@article{Guan2014,
  author  = {Guan, Lei and Zhu, Anding},
  title   = {Green Communications: Digital Predistortion for Wideband {RF} Power Amplifiers},
  journal = {IEEE Microwave Magazine},
  volume  = {15},
  number  = {7},
  pages   = {84--99},
  year    = {2014},
  doi     = {10.1109/MMM.2014.2356037}
}

@article{Leibfried2003,
  title={Experimental demonstration of a robust, high-fidelity geometric two-ion-qubit phase gate},
  author={Leibfried, D. and others},
  journal={Nature},
  volume={422},
  pages={412--415},
  year={2003},
  doi={10.1038/nature01492}
}

@article{Gaebler2016,
  title={High-fidelity universal gate set for {$^{9}$Be$^{+}$} ion qubits},
  author={Gaebler, J. P. and others},
  journal={Physical Review Letters},
  volume={117},
  pages={060505},
  year={2016},
  doi={10.1103/PhysRevLett.117.060505}
}

@article{Hayes2012,
  title={Coherent error suppression in multiqubit entangling gates},
  author={Hayes, D. and others},
  journal={Physical Review Letters},
  volume={109},
  pages={020503},
  year={2012},
  doi={10.1103/PhysRevLett.109.020503}
}

@article{Sapiroid2023,
  title = {Robust Two-Qubit Gates for Trapped Ions Using Spin-Dependent Squeezing},
  author = {Shapira, Yotam and Cohen, Sapir and Akerman, Nitzan and Stern, Ady and Ozeri, Roee},
  journal = {Physical Review Letters},
  volume = {130},
  pages = {030602},
  year = {2023},
  doi = {10.1103/PhysRevLett.130.030602}
}

@misc{SuppMat,
  title = {See Supplemental Material at [URL will be inserted by publisher] for phase-response calibration details, stability characterization, extended spectral benchmarking, and the spectrum-based fidelity estimation procedure.},
  note  = {Supplemental Material to: Digital Predistortion of Optical Fields for Fast and High-Fidelity Entangling Gates in Trapped-Ion Qubits},
  year  = {2026}
}

@article{Ospelkaus2011,
  author  = {Ospelkaus, C. and Langer, C. E. and Amini, J. M. and Brown, K. R. and Leibfried, D. and Wineland, D. J.},
  title   = {Microwave quantum logic gates for trapped ions},
  journal = {Nature},
  volume  = {476},
  pages   = {181--184},
  year    = {2011},
  doi     = {10.1038/nature10290}
}

@article{Zarantonello2019,
  author  = {Zarantonello, G. and Hahn, H. and Morgner, J. and Schulte, M. and Bautista-Salvador, A. and Werner, R. F. and Hammerer, K. and Ospelkaus, C.},
  title   = {Robust and Resource-Efficient Microwave Near-Field Entangling {$^{9}$Be$^{+}$} Gate},
  journal = {Physical Review Letters},
  volume  = {123},
  pages   = {260503},
  year    = {2019},
  doi     = {10.1103/PhysRevLett.123.260503}
}

@article{Weber2024,
  author  = {Weber, M. A. and Gely, M. F. and Hanley, R. K. and Harty, T. P. and Leu, A. D. and L\"{o}schnauer, C. M. and Nadlinger, D. P. and Lucas, D. M.},
  title   = {Robust and fast microwave-driven quantum logic for trapped-ion qubits},
  journal = {Physical Review A},
  volume  = {110},
  pages   = {L010601},
  year    = {2024},
  doi     = {10.1103/PhysRevA.110.L010601}
}

@article{Mintert2001,
  author  = {Mintert, F. and Wunderlich, C.},
  title   = {Ion-Trap Quantum Logic Using Long-Wavelength Radiation},
  journal = {Physical Review Letters},
  volume  = {87},
  pages   = {257904},
  year    = {2001},
  doi     = {10.1103/PhysRevLett.87.257904}
}

@article{Nunnerich2025,
  author  = {N\"{u}nnerich, M. and Cohen, J. and Barthel, T. and Huber, T. and Niroomand, A. and Retzker, A. and Wunderlich, C.},
  title   = {Fast, Robust, and Laser-Free Universal Entangling Gates for Trapped-Ion Quantum Computing},
  journal = {Physical Review X},
  volume  = {15},
  pages   = {021079},
  year    = {2025},
  doi     = {10.1103/PhysRevX.15.021079}
}

@article{Srinivas2021,
  author  = {Srinivas, R. and Burd, S. C. and Knaack, H. M. and Sutherland, R. T. and Kwiatkowski, A. and Glancy, S. and Knill, E. and Wineland, D. J. and Leibfried, D. and Wilson, A. C. and Allcock, D. T. C. and Slichter, D. H.},
  title   = {High-fidelity laser-free universal control of trapped ion qubits},
  journal = {Nature},
  volume  = {597},
  pages   = {209--213},
  year    = {2021},
  doi     = {10.1038/s41586-021-03809-4}
}

@article{Gustavsson2013,
  title   = {Improving Quantum Gate Fidelities by Using a Qubit to Measure Microwave Pulse Distortions},
  author  = {Gustavsson, S. and Zwier, O. and Bylander, J. and Yan, F. and Yoshihara, F. and Nakamura, Y. and Orlando, T. P. and Oliver, W. D.},
  journal = {Physical Review Letters},
  volume  = {110},
  pages   = {040502},
  year    = {2013},
  doi     = {10.1103/PhysRevLett.110.040502}
}

@article{Jerger2019,
  title   = {In Situ Characterization of Qubit Control Lines: A Qubit as a Vector Network Analyzer},
  author  = {Jerger, M. and Kulikov, A. and Vasselin, Z. and Fedorov, A.},
  journal = {Physical Review Letters},
  volume  = {123},
  pages   = {150501},
  year    = {2019},
  doi     = {10.1103/PhysRevLett.123.150501}
}

@article{Rol2020,
  title   = {Time-domain characterization and correction of on-chip distortion of control pulses in a quantum processor},
  author  = {Rol, M. A. and Ciorciaro, L. and Malinowski, F. K. and Tarasinski, B. M. and Sagastizabal, R. E. and Bultink, C. C. and Salath\'{e}, Y. and Haandbaek, N. and Sedivy, J. and DiCarlo, L.},
  journal = {Applied Physics Letters},
  volume  = {116},
  pages   = {054001},
  year    = {2020},
  doi     = {10.1063/1.5133894}
}

@article{Guo2024,
  title   = {Universal Scalable Characterization and Correction of Pulse Distortions in Controlled Quantum Systems},
  author  = {Guo, L.-L. and Duan, P. and Zhang, S. and Yang, X.-X. and Zhang, C. and Du, L. and Zhang, H.-F. and Tao, H.-R. and Wang, T.-L. and Jia, Z.-L. and others},
  journal = {Physical Review Applied},
  volume  = {21},
  pages   = {064060},
  year    = {2024},
  doi     = {10.1103/PhysRevApplied.21.064060}
}

@misc{Hellings2025,
  title   = {Calibrating Magnetic Flux Control in Superconducting Circuits by Compensating Distortions on Time Scales from Nanoseconds up to Tens of Microseconds},
  author  = {Hellings, Christoph and Lacroix, Nathan and Remm, Ants and Boell, Richard and Herrmann, Johannes and Laz\u{a}r, Stefania and Krinner, Sebastian and Swiadek, Fran\c{c}ois and Andersen, Christian Kraglund and Eichler, Christopher and Wallraff, Andreas},
  year    = {2025},
  eprint  = {2503.04610},
  archivePrefix = {arXiv},
  primaryClass  = {quant-ph}
}

@misc{Aggarwal2025,
  title   = {Mitigating transients in flux-control signals in a superconducting quantum processor},
  author  = {Aggarwal, Anuj and Fern\'{a}ndez-Pend\'{a}s, Jorge and Abad, Tahereh and Shiri, Daryoush and Jakobsson, Halld\'{o}r and Rommel, Marcus and Nylander, Andreas and Hogedal, Emil and Osman, Amr and Bizn\'{a}rov\'{a}, Janka and Rehammar, Robert and Faucci Giannelli, Michele and Fadavi Roudsari, Anita and Bylander, Jonas and Tancredi, Giovanna},
  year    = {2025},
  eprint  = {2503.08645},
  archivePrefix = {arXiv},
  primaryClass  = {quant-ph}
}

@article{Ni2025,
  title   = {Correcting on-chip distortion of control pulses with silicon spin qubits},
  author  = {Ni, Ming and Ma, Rong-Long and Kong, Zhen-Zhen and others},
  journal = {Chinese Physics B},
  volume  = {34},
  pages   = {010308},
  year    = {2025},
  doi     = {10.1088/1674-1056/ad8db1}
}

@misc{Duan2026,
  title   = {Measuring and correcting nanosecond pulse distortions in quantum-dot spin qubits},
  author  = {Duan, Jiheng and Torres-Leal, Fernando and Nichol, John M.},
  year    = {2026},
  eprint  = {2602.17899},
  archivePrefix = {arXiv},
  primaryClass  = {quant-ph}
}

@article{Gely2024,
  title   = {In-situ characterization of qubit drive-phase distortions},
  author  = {Gely, M. F. and Litarowicz, J. M. and Leu, A. D. and Lucas, D. M.},
  journal = {Physical Review Applied},
  volume  = {22},
  pages   = {024001},
  year    = {2024},
  doi     = {10.1103/PhysRevApplied.22.024001}
}

@misc{Jaeger2026,
  title   = {Digital Predistortion of Power Amplifiers for Quantum Computing},
  author  = {Jaeger, Marvin and Tegowski, Bartosz and Riemschneider, Georg Frederik and Koelpin, Alexander},
  year    = {2026},
  eprint  = {2601.06524},
  archivePrefix = {arXiv},
  primaryClass  = {quant-ph}
}

\clearpage
\onecolumngrid
\appendix
\setcounter{figure}{0}
\setcounter{table}{0}
\setcounter{equation}{0}
\setcounter{section}{0}
\renewcommand{\thefigure}{S\arabic{figure}}
\renewcommand{\thetable}{S\arabic{table}}
\counterwithout{equation}{section}
\renewcommand{\theequation}{S\arabic{equation}}
\renewcommand{\thesection}{S\arabic{section}}
\renewcommand{\thesubsection}{\thesection.\arabic{subsection}}
\makeatletter
\renewcommand{\p@subsection}{}  
\makeatother
\renewcommand{\theHfigure}{S\arabic{figure}}
\renewcommand{\theHtable}{S\arabic{table}}
\renewcommand{\theHequation}{S\arabic{equation}}
\renewcommand{\theHsection}{S\arabic{section}}
\section*{Supplemental Material}

\section{Phase-response calibration}
\label{sec:supp-phase-calibration}

In addition to the compressive amplitude nonlinearity characterized in the main text, the AOM exhibits amplitude-to-phase conversion: the optical phase of the diffracted beam acquires a systematic shift that depends on the instantaneous RF drive amplitude (\figref{fig:aom_response}{b}). This effect arises from the power dependence of the acoustic velocity due to nonlinear elasticity of the crystal lattice~\cite{Goutzoulis1994, Vernaleken2007}, as well as thermal gradients induced by RF power dissipation that modify the refractive index and acoustic propagation speed~\cite{EsquivelRamirez2024}. Although the present DPD implementation corrects only the amplitude response, an accurate phase calibration is required for the forward simulations used to predict gate performance (Sec.~\ref{sec:supp-simulation}).

\subsection{Heterodyne measurement}

The phase response is measured using the heterodyne setup detailed in Sec.~\ref{sec:supp-heterodyne}: the first-order diffracted beam from the gate AOM is interfered with a reference beam derived from the same 674~nm laser, with the two paths frequency-shifted to produce a $\sim$20~MHz beat on a fast photodiode. The photodiode signal is demodulated in quadrature (I/Q) to extract the instantaneous optical phase of the beat note, from which the AOM-induced phase shift is isolated.

During the measurement, the normalized RF drive amplitude $A$ is stepped from 0 to 1 in fine increments, where $A = 1$ corresponds to the maximum drive amplitude defined in the amplitude calibration (Sec.~\ref{sec:aom-characterization} of the main text). At each amplitude setting, the extracted phase is averaged over multiple repetitions to suppress shot-to-shot noise. The resulting data, $\phi(A)$, are shown in \figref{fig:aom_response}{b}.

\subsection{Polynomial fit and boundary conditions}

The measured phase is modeled by a polynomial of the same form used for the amplitude response,
\begin{equation}
\phi(A) = \sum_{k=1}^{K} b_k A^k,
\label{eq:phase-poly}
\end{equation}
where $b_k$ are real coefficients. The summation begins at $k = 1$ (no constant term), which enforces the physical boundary condition $\phi(0) = 0$: in the absence of RF drive there is no diffracted beam and hence no measurable phase shift.

At low drive amplitudes ($A \lesssim 0.1$), the diffracted beam power is too weak to yield a reliable phase measurement. We therefore fit Eq.~\eqref{eq:phase-poly} to data in the range $0.1 \le A \le 1$ using a fifth-degree polynomial ($K = 5$), and rely on the $\phi(0) = 0$ constraint to smoothly extrapolate the fit to lower amplitudes. The resulting polynomial defines the static phase transfer function used in the AOM forward model (Sec.~\ref{sec:supp-simulation}).

\FloatBarrier

\section{Stability of the amplitude and phase calibrations}
\label{sec:supp-stability}

DPD relies on the assumption that the calibrated transfer functions remain valid between calibration and use. If the AOM response drifts significantly on experimental timescales, the predistortion lookup becomes stale and the correction degrades. We assess the temporal stability of both the amplitude and phase calibrations by repeating the characterization procedures described in Sec.~\ref{sec:aom-characterization} of the main text and Sec.~\ref{sec:supp-phase-calibration} over extended periods.

\subsection{Amplitude stability}

We track the amplitude response by monitoring $y(1)$, the optical amplitude at full RF drive ($A=1$). Because the response is normalized to unit slope at the origin, $y(1)$ equals the maximum-correctable amplitude $A_{\mathrm{corr}}$ (Sec.~\ref{sec:aom-characterization} of the main text). This single scalar summarizes the compressive behavior of the entire transfer curve: a drift in $y(1)$ would indicate that the onset of compression has shifted, invalidating the stored predistortion function.

\figref{fig:stability_timeseries}{a} shows $y(1)$ extracted from $n = 34$ independent calibration runs spanning $\sim$110~days. The mean value is $\langle y(1) \rangle = 0.5655$ with a standard deviation of $\sigma = 0.0091$, corresponding to a relative variation of 1.6\%. The full families of amplitude calibration curves are overlaid in \figref{fig:response_envelopes}{a}, where the narrow $\pm 1\sigma$ envelope confirms that the shape of the transfer function, not merely its compression onset, is reproducible across runs.

\subsection{Phase stability}

\begin{figure}[h!]
\centering
\includegraphics[width=0.9\textwidth]{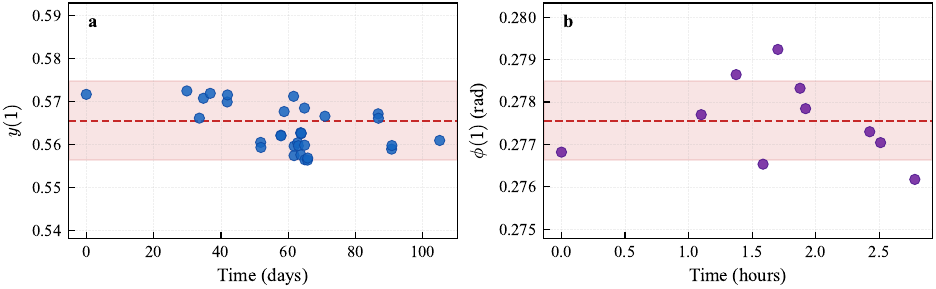}
\caption{Calibration stability over time. (a)~Optical amplitude at full RF drive, $y(1)$, extracted from $n = 34$ repeated calibrations over a 110-day span. Dots: individual runs; dashed line: mean ($0.5655$); shaded band: $\pm 1\sigma$ ($0.0091$). (b)~Phase shift at full RF drive, $\phi(1)$, extracted from $n = 6$ repeated calibrations over a 3-hour session. Dots: individual measurements; dashed line: mean ($0.2776$~rad); shaded band: $\pm 1\sigma$ ($0.0009$~rad).}
\label{fig:stability_timeseries}
\end{figure}

\begin{figure}[h!]
\centering
\includegraphics[width=0.9\textwidth]{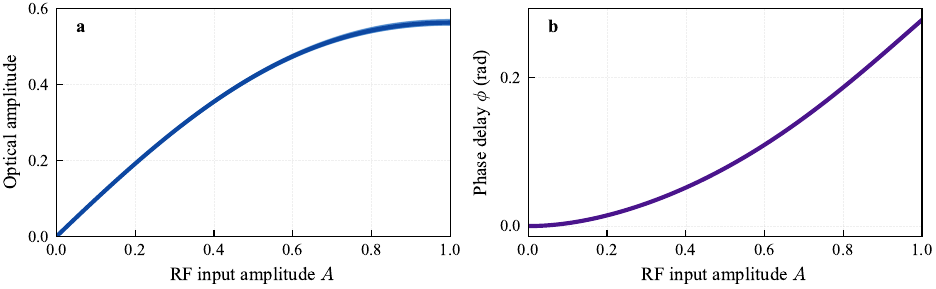}
\caption{Response envelopes for amplitude and phase. (a)~Amplitude response: solid blue line shows the mean of $n=34$ calibration runs; shaded band shows $\pm 1\sigma$. (b)~Phase response: solid purple line shows the mean of $n=6$ calibration runs; shaded band shows $\pm 1\sigma$.}
\label{fig:response_envelopes}
\end{figure}

We track the phase response by monitoring $\phi(1)$, the phase shift at full RF drive ($A=1$). \figref{fig:stability_timeseries}{b} shows $\phi(1)$ extracted from $n = 6$ independent calibration runs spanning $\sim$3~hours. The mean value is $\langle \phi(1) \rangle = 0.2776$~rad with a standard deviation of $\sigma = 0.0009$~rad, corresponding to a relative variation of 0.3\%. The full families of phase calibration curves are overlaid in \figref{fig:response_envelopes}{b}, where the narrow $\pm 1\sigma$ envelope confirms that the shape of the transfer function is reproducible across runs.

The amplitude calibration dataset spans a longer period because this measurement is part of the routine experimental workflow, serving not only DPD but also laser-power stabilization and other AOM housekeeping tasks, and was therefore repeated regularly over the course of normal operation. The phase calibration, by contrast, enters only through the forward simulations (Sec.~\ref{sec:supp-simulation}) and has no role in the experimental control loop, so it was performed in a dedicated session for this study.

\subsection{Practical implications}

The small observed drifts indicate that a single amplitude characterization is sufficient for typical experimental campaigns lasting hours to days. The phase calibration was stable over the $\sim$3-hour measurement session; longer-term phase stability has not been independently verified, but the phase response enters only through the forward simulations (Sec.~\ref{sec:supp-simulation}) and is not used in the DPD correction loop. In our experiments, the amplitude response was recalibrated at the start of each daily session as a precaution; however, the data presented here suggest that this frequency is conservative, and recalibration is strictly necessary only after significant changes to the optical or RF chain, such as realignment of the AOM beam path, replacement of RF components, or AOM servicing.

\FloatBarrier

\section{Cardioid gate waveform construction}
\label{sec:supp-waveform}

The Cardioid(1,2) entangling gate~\cite{Shapira2018} uses two pairs of bichromatic tones placed symmetrically around the qubit carrier frequency $\omega_c$, with each pair detuned from the motional sideband by integer multiples of $\xi_0$. The blue-detuned tones are at $\omega_{b,j} = \omega_c + \nu + j\xi_0$ and the red-detuned tones at $\omega_{r,j} = \omega_c - \nu - j\xi_0$ for $j = 1, 2$, where $\nu$ is the motional-mode frequency and $\xi_0$ is the gate-rate detuning. The gate duration is $T_g = 1/\xi_0$.

In the rotating frame of the carrier, the RF envelope driving the AOM is constructed as
\begin{equation}
  w_I(t) = \sum_{j} r_j \cos\!\bigl(2\pi f_j\,t + \phi_j\bigr),\qquad
  w_Q(t) = \sum_{j} r_j \sin\!\bigl(2\pi f_j\,t + \phi_j\bigr),
  \label{eq:waveform-iq}
\end{equation}
where the sum runs over the four sideband frequencies $f_j \in \{-\nu - 2\xi_0,\; -\nu - \xi_0,\; +\nu + \xi_0,\; +\nu + 2\xi_0\}$, with corresponding relative amplitudes $r_j$ and phases $\phi_j$.

For the Cardioid(1,2) gate the two harmonic indices are $n = 1$ and $n = 2$, with equal amplitudes and a relative $\pi$ phase shift between them: $r_1 = +1$, $r_2 = -1$ (equivalently, $|r_1| = |r_2|$ with $\phi_2 - \phi_1 = \pi$). The red and blue sideband pairs share the same amplitude ratios. The overall waveform is then normalized so that
\begin{equation}
  \max_t \bigl|w_I(t) + i\,w_Q(t)\bigr| = A,
\end{equation}
where $A \in [0, 1]$ is the normalized peak drive amplitude defined in Sec.~\ref{sec:aom-characterization} of the main text. In practice, the AWG output voltage is calibrated so that $A = 1$ corresponds to the RF drive at which the first-order diffraction efficiency reaches its maximum.

When DPD is applied, the inverse amplitude transfer function $f_{\mathrm{AM}}^{-1}$ (Sec.~\ref{sec:supp-simulation}) is applied sample-by-sample to the envelope magnitude, inflating the instantaneous RF amplitude where needed to compensate the AOM's compressive response. Because the symmetric double-sideband encoding places tones at $\pm\nu \pm n\xi_0$, the resulting RF waveform is real-valued ($w_Q = 0$), and the AOM is driven as a single-channel amplitude modulator; the predistortion therefore reduces to a scalar mapping on $w_I(t)$.

The cubic third-order nonlinearity of the AOM generates intermodulation products at frequencies $2f_1 - f_2$ and $2f_2 - f_1$ for each pair of tones. For the Cardioid(1,2) spectrum with tones at detunings $\xi_0$ and $2\xi_0$ from the motional sideband, the dominant IM3 products fall at detunings $0$ (on-resonance, $n = 0$) and $3\xi_0$ ($n = 3$). As discussed in Sec.~\ref{sec:supp-fidelity-estimation}, the on-resonance product is far more damaging to gate fidelity because it produces a phase-space trajectory that fails to close at the gate time.

\FloatBarrier

\section{Heterodyne photodiode spectroscopy}
\label{sec:supp-heterodyne}

The spectral benchmarking measurements reported in the main text rely on heterodyne detection of the gate beam to resolve individual tone powers with high dynamic range. This section describes the optical setup, signal acquisition, and spectral extraction procedure.

\subsection{Optical setup and beat-note generation}

The 674~nm laser source is split into two paths: a gate path and a reference path, each containing a double-pass AOM followed by a single-pass AOM. The gate waveform is programmed on the single-pass AOM of the gate path. The reference path is configured as a single-frequency beam whose carrier is offset by $\sim$20~MHz from the gate beam carrier. A pick-off of the gate beam is combined with the reference beam on a fast photodiode. The resulting heterodyne beat signal on the photodiode encodes the gate waveform spectrum around this intermediate frequency: each gate tone at detuning $n\xi_0$ from the motional sideband produces a beat component at $f_{\mathrm{det}} \pm (\nu + n\xi_0)$, where $f_{\mathrm{det}} \approx 20$~MHz is the detection offset and $\nu$ is the motional-mode frequency.

\subsection{Signal acquisition}

The photodiode signal is digitized by an FPGA-based pulse controller at a sampling rate of 1~GHz. For the spectral benchmarking measurements, the gate waveform is generated with a detuning of $\xi_0 = 20$~kHz (gate time $T_g = 50~\mu$s), larger than the values used for the ion-gate experiments ($\xi_0 \approx 3$--$7$~kHz). The higher detuning shortens the individual gate period, allowing more repetitions to fit within the FPGA acquisition memory and thereby improving the frequency resolution of the recorded spectrum. The total acquisition window is $\sim$2~ms, yielding a frequency resolution of $\Delta f \lesssim 500$~Hz, sufficient to resolve the gate tones (separated by $\xi_0$) from the intermodulation products.

\subsection{Spectral extraction}

The power spectral density (PSD) is computed from the digitized photodiode voltage using a periodogram with a flat-top window, chosen for its amplitude accuracy at the expense of broader spectral leakage.

Tone powers $P_n$ at harmonic indices $n = 0, 1, 2, 3$ are extracted as follows. For the gate tones ($n = 1, 2$), the spectral peak is located within a $\pm 5$~kHz search window around the expected beat frequency, and validated by requiring $< 1$~kHz frequency error and $> 20$~dB signal-to-noise ratio (relative to the median noise floor). For the intermodulation products ($n = 0, 3$), which may be weak or buried in noise, the PSD is sampled directly at the expected frequency without peak searching.

The gate-tone-to-IM3 power ratios $R_{10}$ and $R_{23}$ reported in the main text are computed from the extracted tone powers as $R_{10} = P_{1,\mathrm{dB}} - P_{0,\mathrm{dB}}$ and $R_{23} = P_{2,\mathrm{dB}} - P_{3,\mathrm{dB}}$, where the subscript denotes the harmonic index. Each measurement is performed independently for the blue-detuned ($+\nu$) and red-detuned ($-\nu$) sideband groups; the values reported in the main text are averages over both sidebands.

\FloatBarrier

\section{Gate fidelity estimation from measured spectra}
\label{sec:supp-fidelity-estimation}

The estimated gate fidelity $\mathcal{F}_{\mathrm{est}}$ reported throughout the main text is derived from the phase-space formalism of Ref.~\cite{Shapira2018}. This section presents the formalism, describes how it is applied to photodiode-measured or simulated spectra containing spurious intermodulation (IM) products, and quantifies the relative impact of different IM harmonics on gate performance.

\subsection{Phase-space formalism}

A M\o lmer--S\o rensen entangling gate operates by applying bichromatic laser fields that exert state-dependent forces on the motional mode of an ion crystal.
These forces drive the motional state along a closed trajectory in phase space; the geometric phase accumulated by the enclosed area produces the desired entangling operation.
When the drive spectrum contains tones at harmonic indices $n$ of the gate detuning $\xi_0$, with relative amplitudes $r_n$ and phases $\phi_n$, the motional-mode phase-space coordinates evolve as
\begin{align}
F(t) &= -\sqrt{2}\,\frac{\eta\Omega}{2\pi\xi_0}\sum_{n} \frac{r_n}{n}\sin(2\pi n\xi_0 t + \phi_n), \label{eq:F}\\[4pt]
G(t) &= \sqrt{2}\,\frac{\eta\Omega}{2\pi\xi_0}\sum_{n} \frac{r_n}{n}\bigl[1 - \cos(2\pi n\xi_0 t + \phi_n)\bigr], \label{eq:G}
\end{align}
where the sum runs over all tones present in the drive spectrum, each detuned by $n\xi_0$ from the motional sideband with $n$ a positive integer, $\eta$ is the Lamb--Dicke parameter, and $\Omega$ is the calibrated Rabi frequency.
The geometric phase accumulated by the trajectory is
\begin{equation}
\Phi(t) = -\int_0^t F(\tau)\,\frac{dG}{d\tau}\,d\tau.\label{eq:Phi}
\end{equation}
A maximally entangling gate requires two conditions to be satisfied at the gate time $T_g = 1/\xi_0$: (i)~phase-space closure, $F(T_g) = G(T_g) = 0$, which decouples the spin and motional degrees of freedom; and (ii)~the correct geometric phase, $\Phi(T_g) = \pi/2$, which produces a maximally entangled Bell state.
For the ideal Cardioid(1,2) gate (tones at $n = 1, 2$ only), both conditions are satisfied by construction~\cite{Shapira2018}.

\subsection{Effect of intermodulation products}

The AOM nonlinearity generates spurious tones at harmonic indices beyond the intended gate spectrum, and the impact of these tones on gate fidelity depends critically on whether the harmonic index is zero or a nonzero integer.

At $t = T_g = 1/\xi_0$, any tone with nonzero integer harmonic index $n$ satisfies $2\pi n\xi_0 T_g = 2\pi n$, so that $\sin(2\pi n) = 0$ and $\cos(2\pi n) = 1$. Such tones therefore close exactly in phase space, $F(T_g) = G(T_g) = 0$, and affect only the accumulated geometric phase $\Phi(T_g)$. In particular, the $n = 3$ third-order intermodulation product of the Cardioid(1,2) gate perturbs $\Phi$ but does not cause trajectory non-closure.

The $n = 0$ (on-resonance) tone is qualitatively different. Taking the limit $n \to 0$ in Eqs.~\eqref{eq:F}--\eqref{eq:G},
\begin{equation}
\lim_{n\to0}\frac{\sin(2\pi n\xi_0 t)}{n}=2\pi\xi_0 t,\qquad
\lim_{n\to0}\frac{1-\cos(2\pi n\xi_0 t)}{n}=0,
\end{equation}
the resonant contribution becomes
\begin{equation}
F_0(t)=-\sqrt{2}\,\eta\Omega\,r_0\,t,
\qquad G_0(t)=0,
\end{equation}
a displacement that grows linearly in time and does not return to zero at $T_g$. The resulting non-closure, $F(T_g) \neq 0$, leaves residual spin-motion entanglement and dominates the gate infidelity.

\begin{figure}[h!]
\centering
\includegraphics[width=0.7\textwidth]{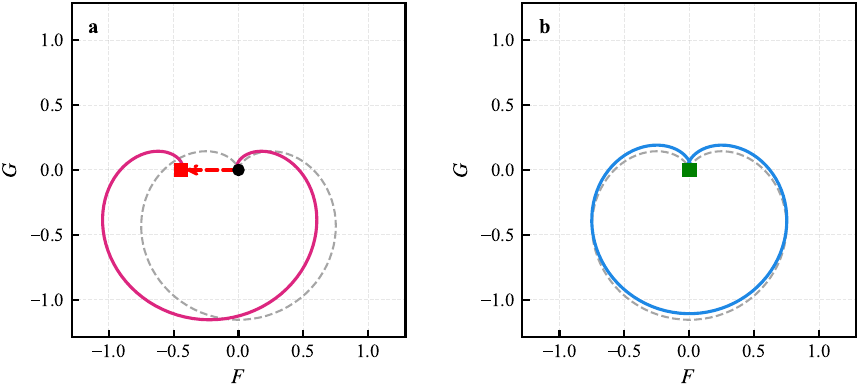}
\caption{Phase-space trajectories for the Cardioid(1,2) gate with a single added IM3 tone at 10\% relative amplitude. Gray dashed line: ideal trajectory; black dot: origin. (a)~$n = 0$ resonant tone (magenta): the trajectory fails to close, ending at $F(T_g) \approx -0.44$ (red square). (b)~$n = 3$ integer-harmonic tone (blue): the trajectory returns to the origin (green square), with only a small perturbation to the geometric phase.}
\label{fig:phase_space_n0_vs_n3}
\end{figure}

Figure~\ref{fig:phase_space_n0_vs_n3} illustrates this contrast. An $n = 0$ tone at 10\% of the gate-tone amplitude produces a large phase-space offset (\figref{fig:phase_space_n0_vs_n3}{a}), whereas the same amplitude in an $n = 3$ tone yields a trajectory that closes exactly (\figref{fig:phase_space_n0_vs_n3}{b}).

\subsection{Rabi-frequency calibration and gate fidelity}

In practice, the Rabi frequency $\Omega$ is not a free parameter but is determined by the optical power in the gate tones. Calibrating $\Omega$ from only the designed tones ($n = 1, 2$) so that $\Phi(T_g) = \pi/2$ in the absence of intermodulation contamination yields
\begin{equation}
\Omega=\frac{\pi\xi_0}{\eta}\biggl(\sum_{n\in\mathrm{gate}}\frac{r_n^2}{n}\biggr)^{-1/2}.\label{eq:Omega_cal}
\end{equation}
For the Cardioid(1,2) gate with amplitudes $r_1 = +1/\sqrt{1.5}$, $r_2 = -1/\sqrt{1.5}$ (equal magnitudes, $\pi$ relative phase, matching the sign convention of Sec.~\ref{sec:supp-waveform}), normalized such that $\sum r_n^2/n = 1$, Eq.~\eqref{eq:Omega_cal} reduces to $\Omega = \pi\xi_0/\eta$. This calibrated value of $\Omega$ is then used in Eqs.~\eqref{eq:F}--\eqref{eq:Phi} to evaluate the phase-space trajectory for the full drive spectrum, including any spurious intermodulation tones. Any deviation of $F(T_g)$, $G(T_g)$, or $\Phi(T_g)$ from their ideal values is therefore attributable to the spectral contamination.

The resulting Bell-state fidelity for a M\o lmer--S\o rensen gate acting on a thermal motional state with mean phonon number $\bar{n}$ is~\cite{Shapira2018}
\begin{equation}
\mathcal{F}=\frac{3+e^{-4(\bar{n}+1/2)(F^2+G^2)}}{8}+\frac{e^{-(\bar{n}+1/2)(F^2+G^2)}}{2}\sin^2\!\Bigl(\Phi+\frac{FG}{2}\Bigr),\label{eq:fidelity}
\end{equation}
where $F \equiv F(T_g)$, $G \equiv G(T_g)$, and $\Phi \equiv \Phi(T_g)$. Non-closure ($F, G \neq 0$) exponentially suppresses the coherent term, while geometric-phase errors shift the argument of $\sin^2$ away from $\pi/2$. For the results in the main text we use $\bar{n} = 0.1$, consistent with sideband thermometry of the stretch mode (Sec.~\ref{sec:experimental-setup} of the main text).

\subsection{Application to measured spectra}

To estimate the gate fidelity from a heterodyne photodiode measurement or a forward simulation, we:
\begin{enumerate}
  \item Convert the measured tone powers $P_n$ (in dB) to linear amplitudes $|r_n| = \sqrt{10^{P_n/10}}$ and normalize to the $n = 1$ gate tone ($|r_1| \to 1$). Assign signs from the Cardioid(1,2) design: $r_1 > 0$, $r_2 < 0$ (Sec.~\ref{sec:supp-waveform}).
  \item Calibrate $\Omega$ via Eq.~\eqref{eq:Omega_cal} using only the gate-tone amplitudes ($n = 1, 2$).
  \item Evaluate $F(T_g)$, $G(T_g)$, and $\Phi(T_g)$ from Eqs.~\eqref{eq:F}--\eqref{eq:Phi} using the full spectrum, including the IM products at $n = 0$ and $n = 3$.
  \item Compute $\mathcal{F}_{\mathrm{est}}$ from Eq.~\eqref{eq:fidelity}.
\end{enumerate}
IM tone phases are assigned from the cubic nonlinearity model: for Cardioid(1,2) with signed gate-tone amplitudes $r_1 > 0$, $r_2 < 0$, the $n = 0$ product has amplitude $\propto r_1^2 r_2 < 0$ (phase~$\pi$) while the $n = 3$ product has amplitude $\propto r_2^2 r_1 > 0$ (phase~0). Gate-tone phases are encoded in the sign of the amplitudes $r_n$.
This procedure is applied independently to the blue-detuned and red-detuned sideband groups; the reported $\mathcal{F}_{\mathrm{est}}$ in the main text is the average over both sidebands.

The key physical insight is that the $n = 0$ resonant tone drives $F(T_g) \propto r_0\, T_g$, so even a small IM amplitude produces a large infidelity through trajectory non-closure. By contrast, the $n = 3$ integer-harmonic tone closes in phase space and contributes only a perturbation to the geometric phase.

\subsection{IM3 suppression requirements}

\begin{figure}[h!]
\centering
\includegraphics[width=0.45\textwidth]{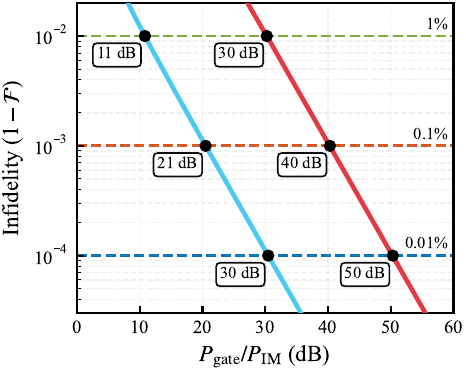}
\caption{Gate infidelity ($1-\mathcal{F}$) versus gate-tone-to-IM power ratio $P_{\mathrm{gate}}/P_{\mathrm{IM}}$ for the Cardioid(1,2) gate. Left curve: resonant $n = 0$ intermodulation product; right curve: integer-harmonic $n = 3$ product. Horizontal lines mark infidelity thresholds of $10^{-2}$, $10^{-3}$, and $10^{-4}$. Labeled dots indicate the required $P_{\mathrm{gate}}/P_{\mathrm{IM}}$ (dB) at each threshold crossing (see Table~\ref{tab:im_suppression}).}
\label{fig:infidelity_vs_im3_power}
\end{figure}

Figure~\ref{fig:infidelity_vs_im3_power} quantifies this asymmetry by plotting the gate infidelity as a function of the gate-tone-to-IM power ratio $P_{\mathrm{gate}}/P_{\mathrm{IM}}$ for a single spurious tone added to the ideal Cardioid(1,2) spectrum.

Table~\ref{tab:im_suppression} lists the gate-tone-to-IM power ratios at which each infidelity threshold is crossed.
At every threshold, the $n = 0$ tone requires $\sim$20~dB more suppression than $n = 3$. For example, reaching $10^{-3}$ infidelity demands a gate/IM ratio of $\sim$40~dB for the $n = 0$ product but only $\sim$21~dB for the $n = 3$ product. These numbers establish the resonant IM3 product as the critical target for predistortion and motivate the use of $R_{10} = P_{1,\mathrm{dB}} - P_{0,\mathrm{dB}}$ as the primary spectral metric in the main text.

\begin{table}[h!]
\centering
\caption{Required gate-tone-to-IM power ratio $P_{\mathrm{gate}}/P_{\mathrm{IM}}$ (dB) for target infidelity thresholds. The $n = 0$ tone requires $\sim$20~dB more suppression than $n = 3$ at every threshold.}
\label{tab:im_suppression}
\begin{tabular}{ccc}
\hline
Infidelity & $n=0$ (dB) & $n=3$ (dB) \\ \hline
$10^{-2}$  & $30.2$     & $10.8$     \\
$10^{-3}$  & $40.3$     & $20.5$     \\
$10^{-4}$  & $50.3$     & $30.5$     \\ \hline
\end{tabular}
\end{table}

\FloatBarrier

\section{AOM forward simulation and derived axes}
\label{sec:supp-simulation}

The photodiode-derived fidelity predictions shown in the main-text figures are obtained by numerically propagating gate waveforms through a calibrated AOM model. Both simulated and experimental data share the normalized RF amplitude $A$ as their common control parameter; this section describes the forward model, the procedure for converting $A$ into the time-averaged diffraction efficiency $\bar{\eta}$ (used for both simulated and measured data), and the fitting procedure that maps the simulated gate-tone power onto the experimentally measured gate rate $\xi_0$.

\subsection{AOM forward model}

We model the AOM as a memoryless nonlinear system characterized by the two static transfer functions obtained from calibration:
\begin{enumerate}
  \item \emph{Amplitude response}: the optical field-amplitude output $y = f_{\mathrm{AM}}(|x|)$, evaluated from the $K = 8$ polynomial fit to the measured diffraction curve (Eq.~\eqref{eq:amplitude-poly} of the main text) with odd symmetry, $f_{\mathrm{AM}}(-x) = -f_{\mathrm{AM}}(x)$, and the input magnitude clipped to $|x| \le 1$.
  \item \emph{Phase response}: the amplitude-dependent phase shift $\phi = f_{\mathrm{PM}}(|x|)$, evaluated from the $K = 5$ polynomial fit described in Sec.~\ref{sec:supp-phase-calibration} with even symmetry (phase depends only on magnitude).
\end{enumerate}
For a complex input envelope $x(t) = |x(t)|\,e^{i\theta(t)}$, the modeled optical output is
\begin{equation}
  y(t) = f_{\mathrm{AM}}\!\bigl(|x(t)|\bigr)\;\exp\!\bigl[i\bigl(\theta(t) + f_{\mathrm{PM}}\!\bigl(|x(t)|\bigr)\bigr)\bigr].
  \label{eq:aom-forward}
\end{equation}
The simulations used throughout the main text incorporate both the amplitude and phase responses.

\subsection{Simulation procedure}

The simulation takes as input a normalized RF drive amplitude $A$ and a DPD flag. A Cardioid(1,2) gate waveform is constructed as described in Sec.~\ref{sec:supp-waveform}; when DPD is enabled, the sample-by-sample inverse-amplitude correction described in the same section is applied. The resulting RF waveform is then propagated through the forward model (Eq.~\eqref{eq:aom-forward}) to obtain the optical output, whose power spectrum is computed following the heterodyne extraction procedure of Sec.~\ref{sec:supp-heterodyne}. The extracted tone powers $P_n$ at harmonic indices $n = 0, 1, 2, 3$ are passed to the fidelity estimator of Sec.~\ref{sec:supp-fidelity-estimation} to obtain $\mathcal{F}_{\mathrm{est}}(A, \mathrm{DPD})$. Sweeping $A$ over its full range produces the simulation curves shown alongside the photodiode measurements in the main text.

\subsection{Mapping RF amplitude to gate rate}
\label{sec:supp-gate-rate-mapping}

The gate rate is proportional to the Rabi frequency, which scales as the square root of the optical power in the gate tone:
\begin{equation}
  r_{\mathrm{rel}}(A, \mathrm{DPD}) = \sqrt{\frac{P_1(A, \mathrm{DPD})}{P_1^{\mathrm{anchor}}}}
    = 10^{(P_{1,\mathrm{dB}} - P_{1,\mathrm{dB}}^{\mathrm{anchor}})/20},
  \label{eq:r-rel}
\end{equation}
where $P_1$ is the simulated gate-tone ($n = 1$) power and the anchor is the lowest measured setting, $(A = 0.4,\;\text{no DPD})$. This yields a dimensionless relative gate rate with $r_{\mathrm{rel}} = 1$ at the anchor.

To convert to an absolute gate rate, we introduce a scale factor $\alpha$:
\begin{equation}
  \xi_0 = \alpha \cdot r_{\mathrm{rel}}.
  \label{eq:xi0-alpha}
\end{equation}

\subsection{Two-parameter fit for the gate-rate axis}
\label{sec:supp-two-param-fit}

To overlay the photodiode-derived fidelity curves onto the measured gate data (Fig.~\ref{fig:fidelity_vs_gate_rate}), we fit two parameters shared across all curves:
\begin{enumerate}
  \item A horizontal scale factor $\alpha$ converting $r_{\mathrm{rel}}$ to $\xi_0$ (Eq.~\eqref{eq:xi0-alpha}).
  \item A vertical fidelity offset $\delta$ applied to all photodiode curves: $\mathcal{F}_{\mathrm{shifted}} = \mathcal{F}_{\mathrm{PD}} + \delta$.
\end{enumerate}
The offset $\delta$ accounts for error sources not captured by the spectral model (e.g., motional heating, laser phase noise, SPAM; see Sec.~\ref{sec:supp-error-budget}).
Both parameters are determined by weighted least-squares minimization of a stacked residual vector containing horizontal residuals (measured $\xi_0$ versus $\alpha\cdot r_{\mathrm{rel}}$, weighted by combined gate-rate and power-fluctuation uncertainties) and vertical residuals (measured fidelity versus $\mathcal{F}_{\mathrm{PD}} + \delta$, weighted by fidelity error bars).
The fit yields $\alpha = 3.42 \pm 0.06$~kHz and $\delta = -0.018 \pm 0.003$.

\subsection{Gate-rate uncertainty from power fluctuations}
\label{sec:supp-power-fluctuations}

Laser-power fluctuations introduce gate-rate uncertainty. We quantify this by comparing repeated photodiode power measurements against the simulation curve, extracting a power spread of $\sigma_{\mathrm{dB}} \approx 0.2$~dB. Since $\xi_0 \propto \sqrt{P}$, the fractional rate uncertainty is
\begin{equation}
  \frac{\sigma_{\xi_0}}{\xi_0} = \frac{\ln 10}{20}\;\sigma_{\mathrm{dB}},
  \label{eq:sigma-xi}
\end{equation}
giving $\sigma_{\xi_0}/\xi_0 \approx 2.3\%$. Because this is a relative uncertainty, the absolute gate-rate spread $\sigma_{\xi_0} = \alpha\cdot\sigma_{\mathrm{rel}}$ grows linearly with gate rate, producing the widening shaded bands at higher $\xi_0$ in Fig.~\ref{fig:fidelity_vs_gate_rate}.

\subsection{Time-averaged diffraction efficiency}
\label{sec:supp-eta-bar}

The time-averaged diffraction efficiency provides a hardware-independent horizontal axis for Figs.~\ref{fig:fidelity_vs_eff} and~\ref{fig:pd_fidelity_vs_eff}. For a gate waveform with $N$ samples $\{w_i\}$ (after optional DPD), we define
\begin{equation}
  \bar{\eta}(A, \mathrm{DPD}) = \eta_{\mathrm{ref}}\;\frac{\bar{P}_{\mathrm{gate}}}{P_{\mathrm{ref}}},
  \label{eq:eta-bar}
\end{equation}
where
\begin{equation}
  \bar{P}_{\mathrm{gate}} = \frac{1}{N}\sum_{i=1}^{N}\bigl|f_{\mathrm{AM}}(w_i)\bigr|^2
\end{equation}
is the time-averaged optical power at the AOM output,
$P_{\mathrm{ref}} = |f_{\mathrm{AM}}(1)|^2$
is the optical power at full RF drive ($A = 1$), and $\eta_{\mathrm{ref}} = 0.80$ is the measured peak first-order diffraction efficiency at full drive. Here $f_{\mathrm{AM}}$ is the $K = 8$ polynomial amplitude transfer function defined in the main text (Eq.~\eqref{eq:amplitude-poly}). To assign $\bar{\eta}$ to experimental data points at arbitrary $A$, we evaluate $\bar{\eta}(A, \mathrm{DPD})$ on a dense grid of 100 amplitudes spanning the operating range and use cubic spline interpolation between the grid points.

The correctable diffraction efficiency $\bar{\eta}_{\mathrm{corr}}$ is obtained by evaluating $\bar{\eta}$ at $A = A_{\mathrm{corr}}$ for both DPD and no-DPD cases. Because DPD reshapes the RF envelope, the same correctable amplitude yields a higher $\bar{\eta}$ after predistortion ($\bar{\eta}_{\mathrm{corr}}^{\mathrm{DPD}} > \bar{\eta}_{\mathrm{corr}}^{\mathrm{NoDPD}}$), extending the usable operating range to higher optical throughput.

\FloatBarrier

\section{Gate fidelity extraction}
\label{sec:supp-fidelity-extraction}

The Bell-state fidelity reported in the main text is obtained from two independent measurements at the gate time $T_g$: a population measurement and a parity-fringe scan~\cite{Sackett2000}. At each RF amplitude setting, multiple independent scans were recorded with DPD-enabled and DPD-disabled conditions interlaced within each scan to mitigate slow drifts.

\subsection{Population fidelity}

A projective measurement in the computational basis yields the two-ion state populations $P_{00}$, $P_{01}+P_{10}$, and $P_{11}$ from $N = 625$ experimental repetitions (shots) per scan. The population fidelity is the even-parity fraction,
\begin{equation}
F_{\mathrm{pop}} = P_{00} + P_{11},
\end{equation}
with a binomial uncertainty $\sigma_{\mathrm{pop}} = \sqrt{F_{\mathrm{pop}}(1-F_{\mathrm{pop}})/N}$. Note that $F_{\mathrm{pop}}$ is a single binomial variable (even vs.\ odd parity outcome), not the sum of two independent fractions, so the standard binomial expression applies directly.

\subsection{Parity fidelity}

A parity-fringe scan is performed by applying a variable analysis phase $\varphi$ after the gate and measuring the parity $\Pi(\varphi) = P_{00} + P_{11} - (P_{01}+P_{10})$ at 16 uniformly spaced phases over $[0, 2\pi)$, with $N = 625$ shots per phase point. The data are fitted to a sinusoidal model,
\begin{equation}
\Pi(\varphi) = \mathcal{A}\cos(2\varphi + \varphi_0),
\label{eq:parity-model}
\end{equation}
using maximum-likelihood estimation (MLE) with a binomial likelihood. Specifically, at each phase point $\varphi_i$ the probability of an even-parity outcome is modeled as $p_i = \tfrac{1}{2}[1 + \mathcal{A}\cos(2\varphi_i + \varphi_0)]$, and the negative log-likelihood
\begin{equation}
-\ln\mathcal{L} = -\sum_i \bigl[k_i \ln p_i + (n_i - k_i)\ln(1 - p_i)\bigr]
\end{equation}
is minimized over the amplitude $\mathcal{A}$ and phase offset $\varphi_0$, where $k_i$ and $n_i$ are the even-parity count and total count at $\varphi_i$, respectively. The oscillation frequency is fixed at $2\varphi$ (two-ion Bell state). The parity fidelity is $F_{\mathrm{par}} = |\mathcal{A}|$, and its uncertainty $\sigma_{\mathrm{par}}$ is obtained from the diagonal element of the inverse Hessian of $-\ln\mathcal{L}$ evaluated at the MLE optimum, $\sigma_{\mathrm{par}} = \sqrt{[\mathbf{H}^{-1}]_{\mathcal{A}\mathcal{A}}}$. The Hessian is computed numerically using central finite differences with step size $\epsilon = 10^{-5}$, and its positive definiteness is verified via an eigenvalue check before inverting.

\subsection{Combined fidelity}

The Bell-state fidelity is
\begin{equation}
\mathcal{F} = \frac{F_{\mathrm{pop}} + F_{\mathrm{par}}}{2},
\end{equation}
with uncertainty $\sigma_{\mathcal{F}} = \tfrac{1}{2}\sqrt{\sigma_{\mathrm{pop}}^2 + \sigma_{\mathrm{par}}^2}$, treating the two error sources as independent.

\subsection{Aggregation across repeated scans}

At each RF amplitude setting, multiple independent scans were recorded. Population measurements were aggregated by count-level pooling: the even-parity counts ($k = \mathrm{round}(F_{\mathrm{pop}} \cdot N)$) and total counts $N$ from all scans at the same amplitude and DPD setting are summed before recomputing $F_{\mathrm{pop}}$ and its binomial error from the pooled totals. For parity measurements, the raw phase-binned counts from repeated scans are merged (phases are rounded to a common 0.01-radian grid and counts at matching phases are summed), and the MLE fit (Eq.~\eqref{eq:parity-model}) is performed on the merged dataset. This count-level aggregation correctly accounts for varying numbers of shots across scans and avoids the information loss inherent in averaging pre-fitted contrasts.

\FloatBarrier

\section{Extended spectral benchmarking: power ratio $R_{23}$}
\label{sec:supp-r23}

The main text quantifies DPD performance through the gate-tone-to-IM3 power ratio $R_{10} = P_{1,\mathrm{dB}} - P_{0,\mathrm{dB}}$, which compares the first gate tone ($n = 1$) to the on-resonance IM3 product ($n = 0$). As shown in Sec.~\ref{sec:supp-fidelity-estimation}, this product dominates the gate infidelity because it causes phase-space non-closure that grows linearly with gate time.

However, the cubic nonlinearity of the AOM generates a second IM3 product at harmonic index $n = 3$ (detuning $3\xi_0$), arising from the mixing term $2f_2 - f_1$ for each sideband pair in the Cardioid(1,2) waveform (Sec.~\ref{sec:supp-waveform}). Although this product closes in phase space and contributes only a geometric-phase perturbation (Sec.~\ref{sec:supp-fidelity-estimation}, Table~\ref{tab:im_suppression}), tracking it provides a more complete picture of how DPD reshapes the output spectrum and verifies that the linearization is not achieved at the expense of other spectral components.

\subsection{Definition}

We define the complementary power ratio
\begin{equation}
R_{23} = P_{2,\mathrm{dB}} - P_{3,\mathrm{dB}},
\end{equation}
which compares the second gate tone ($n = 2$, at detuning $2\xi_0$) to the $n = 3$ IM3 product. Both $R_{10}$ and $R_{23}$ are extracted from the heterodyne photodiode spectra using the procedure described in Sec.~\ref{sec:supp-heterodyne}.

\subsection{Power ratios versus drive amplitude}

Figure~\hyperref[fig:supp_r23]{\ref*{fig:supp_r23}(a,\,b)} shows $R_{10}$ and $R_{23}$ versus normalized RF amplitude $A$. Panel~(a) reproduces the $R_{10}$ data shown in Fig.~\ref{fig:cardioid_spectrum_comparison} of the main text for ease of comparison. Within the correctable range ($A \lesssim A_{\mathrm{corr}}$), DPD improves both power ratios by several dB relative to uncorrected operation. Beyond $A_{\mathrm{corr}}$, the inverse amplitude mapping $f_{\mathrm{AM}}^{-1}$ (Sec.~\ref{sec:supp-simulation}) is clipped at full RF drive and can no longer compensate the AOM compression, so the DPD advantage diminishes. The similar improvement in both $R_{10}$ and $R_{23}$ confirms that DPD suppresses both third-order intermodulation products ($n = 0$ and $n = 3$), not only the single most harmful one.

\subsection{Gate-tone power redistribution}

By linearizing the AOM transfer function, DPD also modifies the absolute optical power delivered to the intended gate tones. We quantify this through the gate-tone power change
\begin{equation}
\Delta P_n = 10\log_{10}\!\left(\frac{P_n^{\mathrm{DPD}}}{P_n^{\mathrm{NoDPD}}}\right),\qquad n \in \{1,\,2\},
\end{equation}
evaluated at matched normalized RF amplitude $A$. Figure~\hyperref[fig:supp_r23]{\ref*{fig:supp_r23}(c)} shows $\Delta P_n$ versus $A$ for both gate tones. The two tones track each other closely, as expected from their equal design amplitudes ($|r_1| = |r_2| = 1/\sqrt{1.5}$; Sec.~\ref{sec:supp-waveform}). The positive $\Delta P_n$ at intermediate amplitudes indicates that DPD recovers gate-tone power that would otherwise be lost to gain compression, channeling it back into the intended spectral components. This power recovery is the physical origin of the effective gate-rate increase discussed in Sec.~\ref{sec:supp-fixed-amplitude}.

\subsection{Comparison with forward simulations}

\begin{figure}[h!]
\centering
\includegraphics[width=\textwidth]{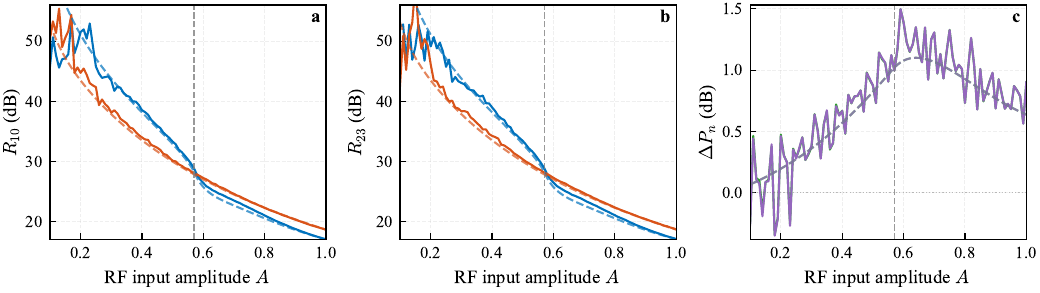}
\caption{Spectral diagnostics of amplitude-only DPD for the Cardioid(1,2) gate waveform as a function of normalized RF drive amplitude $A$. Solid lines: heterodyne photodiode measurements; dashed lines: AOM forward-model simulations incorporating both the amplitude and phase transfer functions. Blue: DPD enabled; red: no DPD. The vertical dashed gray line in each panel marks the correctable amplitude $A_{\mathrm{corr}}$.
(a)~Gate-tone-to-IM3 power ratio $R_{10} = P_{1,\mathrm{dB}} - P_{0,\mathrm{dB}}$, comparing the first gate tone ($n = 1$) to the on-resonance IM3 product ($n = 0$). This panel reproduces the data shown in Fig.~\ref{fig:cardioid_spectrum_comparison} of the main text.
(b)~Complementary power ratio $R_{23} = P_{2,\mathrm{dB}} - P_{3,\mathrm{dB}}$, comparing the second gate tone ($n = 2$) to the $n = 3$ IM3 product.
(c)~Gate-tone power change $\Delta P_n = 10\log_{10}(P_n^{\mathrm{DPD}}/P_n^{\mathrm{NoDPD}})$ for $n{=}1$ (green) and $n{=}2$ (purple), quantifying the optical power recovered by DPD at each gate tone.
Within $A \lesssim A_{\mathrm{corr}}$, DPD improves both power ratios by several dB and recovers gate-tone power lost to compression; beyond $A_{\mathrm{corr}}$ the predistortion runs out of RF headroom and the improvement diminishes.}
\label{fig:supp_r23}
\end{figure}

Dashed lines in all three panels of Fig.~\ref{fig:supp_r23} show the corresponding quantities computed from the AOM forward model (Sec.~\ref{sec:supp-simulation}), which propagates the Cardioid(1,2) waveform through both the calibrated amplitude and phase transfer functions (Sec.~\ref{sec:aom-characterization} of the main text; Sec.~\ref{sec:supp-phase-calibration}). The close agreement between simulation and measurement across all three diagnostics, for both DPD-enabled and uncorrected operation, confirms that the static polynomial model of the AOM nonlinearity is sufficient to capture the observed spectral distortion and the improvement provided by amplitude-only predistortion.

\FloatBarrier

\section{Predistortion effects at fixed target amplitude}
\label{sec:supp-fixed-amplitude}

The main-text gate fidelity measurements (Fig.~\ref{fig:fidelity_vs_gate_rate}) compare DPD and no-DPD performance at matched gate rates, thereby isolating the effect of spectral distortion from that of gate speed. Here we present a complementary comparison at a fixed target amplitude $A = 0.5$, which provides direct insight into the two distinct mechanisms by which DPD improves gate performance: suppression of spurious intermodulation tones and recovery of gate-tone optical power.

An important distinction applies: the predistortion algorithm passes each sample of the target waveform through the inverse amplitude transfer function $f_{\mathrm{AM}}^{-1}$ (Sec.~\ref{sec:supp-simulation}), inflating the instantaneous RF drive to compensate for the AOM compression. As a result, the peak RF amplitude of the corrected waveform exceeds $0.5$; the label $A = 0.5$ refers to the common target modulation amplitude before predistortion, not to the actual applied RF power.

\subsection{Time-resolved two-qubit dynamics}

\begin{figure}[h!]
    \centering
    \includegraphics[width=0.9\textwidth]{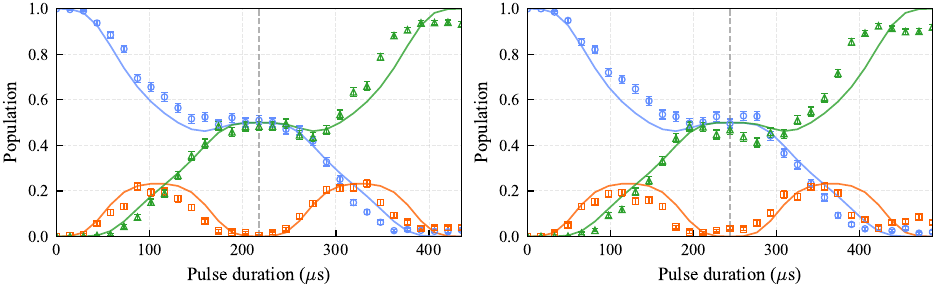}
    \caption{Time-resolved two-ion state populations during the Cardioid(1,2) entangling gate at target amplitude $A = 0.5$. Each point is the mean of $N = 625$ projective measurements; error bars denote $1\sigma$ binomial uncertainties. Hollow markers distinguish the three two-qubit populations: circles (blue) $P_{00}$, squares (orange) $P_{01}{+}P_{10}$, and triangles (green) $P_{11}$. Solid curves are the ideal Cardioid model evaluated at the calibrated gate rate with zero mean phonon number ($\bar{n} = 0$); deviations at longer times reflect dephasing processes not included in the model. The vertical dashed line marks the gate time $T_g = 1/\xi_0$. (Left)~With DPD, $\xi_0 = 4.6$~kHz ($T_g \approx 217~\mu$s): the residual population $P(01{+}10)$ is strongly suppressed at $T_g$, consistent with accurate phase-space closure and spin-motion decoupling. (Right)~Without DPD, $\xi_0 = 4.1$~kHz ($T_g \approx 244~\mu$s): significant $P(01{+}10)$ persists at $T_g$, signaling incomplete closure of the motional phase-space loop due to the on-resonance IM3 product ($n = 0$).}
    \label{fig:timescan-A05}
\end{figure}

Figure~\ref{fig:timescan-A05} shows the time-resolved two-ion state populations $P_{00}$, $P_{01}+P_{10}$, and $P_{11}$ during the Cardioid(1,2) gate at $A = 0.5$. With DPD (left panel), the intermediate-state population $P(01{+}10)$ is strongly suppressed at the gate time $T_g$, indicating that the motional phase-space trajectory closes and the spin and motional degrees of freedom decouple as required for an ideal entangling gate. Without DPD (right panel), a significant residual $P(01{+}10)$ persists at $T_g$. As discussed in Sec.~\ref{sec:supp-fidelity-estimation}, this behavior is the signature of phase-space non-closure caused by the on-resonance IM3 product ($n = 0$), whose contribution to the displacement $F(T_g)$ grows linearly with gate time and does not vanish at $T_g$.

\subsection{Parity-fringe measurements and gate fidelity}

\begin{figure}[h!]
    \centering
    \includegraphics[width=0.9\textwidth]{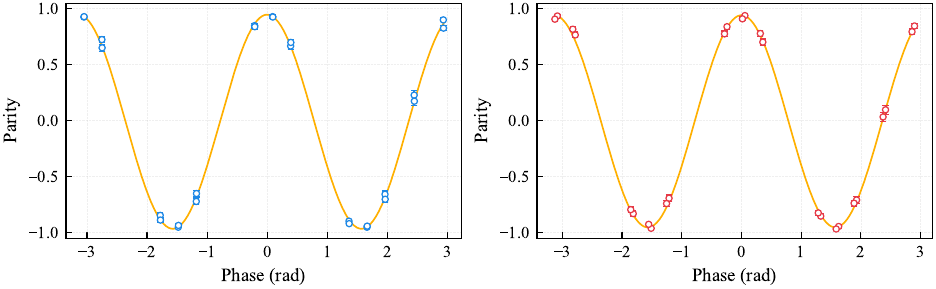}
    \caption{Parity-fringe scans at target amplitude $A = 0.5$. Open markers show the measured parity $\Pi(\varphi)$ as a function of the analysis phase $\varphi$, with $N = 625$ projective measurements per phase point; error bars denote $1\sigma$ binomial uncertainties. Solid curves are maximum-likelihood sinusoidal fits $\Pi(\varphi) = \mathcal{A}\cos(2\varphi + \varphi_0)$ (Eq.~\eqref{eq:parity-model}), where the parity contrast determines the fidelity contribution $F_{\mathrm{par}} = |\mathcal{A}|$ (Sec.~\ref{sec:supp-fidelity-extraction}). Blue (left): DPD-corrected gate, $\mathcal{F}_{\mathrm{DPD}} = 0.9697 \pm 0.0014$. Red (right): uncorrected gate, $\mathcal{F}_{\mathrm{NoDPD}} = 0.9542 \pm 0.0058$. All uncertainties are derived from the Hessian of the binomial log-likelihood at the MLE optimum. The visibly reduced parity contrast without DPD is consistent with residual spin-motion entanglement from incomplete phase-space closure at $T_g$.}
    \label{fig:parity-A05}
\end{figure}

The Bell-state fidelity at this operating point is extracted from parity-fringe scans (Fig.~\ref{fig:parity-A05}) using the MLE procedure described in Sec.~\ref{sec:supp-fidelity-extraction}. The resulting fidelities are
\[
\mathcal{F}_{\mathrm{DPD}}(A = 0.5) = 0.9697 \pm 0.0014, \qquad
\mathcal{F}_{\mathrm{NoDPD}}(A = 0.5) = 0.9542 \pm 0.0058,
\]
a 1.6 percentage-point improvement with DPD. The higher parity contrast in the DPD case (Fig.~\ref{fig:parity-A05}, left) is consistent with improved phase-space closure reducing residual spin-motion entanglement at $T_g$.

\subsection{Effective gate-rate shift}

In addition to suppressing intermodulation products, DPD channels a larger fraction of the optical power into the intended gate tones (as quantified by $\Delta P_n > 0$ in Fig.~\ref{fig:supp_r23}). Because the Rabi frequency scales as $\Omega \propto \sqrt{P_{\mathrm{gate}}}$, this power recovery increases the effective gate rate. For the data presented here, the extracted gate rates are
\[
\xi_0^{\mathrm{DPD}} = 4.6~\mathrm{kHz}, \qquad
\xi_0^{\mathrm{NoDPD}} = 4.1~\mathrm{kHz},
\]
an $\sim$12\% increase at the same target amplitude. This gate-rate shift is a direct consequence of the AOM operating in its compressive regime at $A = 0.5$: the predistortion boosts the RF drive to restore the optical power that would otherwise be lost to compression.

The observed fidelity improvement at fixed $A$ therefore reflects two superimposed effects: (i)~suppression of IM3 distortion, which improves spectral purity, and (ii)~an increase in gate rate, which shortens the gate duration and reduces sensitivity to decoherence. Disentangling these two contributions requires the matched-rate comparison presented in the main text (Fig.~\ref{fig:fidelity_vs_gate_rate}), where the fidelity improvement at equal $\xi_0$ is attributable solely to IM3 suppression.

\FloatBarrier

\section{Error sources and limitations}
\label{sec:supp-error-budget}

The two-parameter fit (Sec.~\ref{sec:supp-two-param-fit}) yields a fidelity offset $\delta = -0.018 \pm 0.003$, indicating that measured gate fidelities are systematically $\sim$1.8\% below the photodiode-derived estimates $\mathcal{F}_{\mathrm{est}}$. Because $\delta$ is a single constant shared across all drive amplitudes and DPD settings, it captures error sources that are approximately independent of the IM3 distortion level. Below we identify and quantify the dominant contributions.

\subsection{State preparation and measurement}
\label{sec:supp-spam}

The reported fidelities are not corrected for state-preparation-and-measurement (SPAM) errors. State detection is performed via state-selective fluorescence on the 422~nm $S_{1/2} \leftrightarrow P_{1/2}$ transition, collected by an EMCCD camera. Camera-based discrimination of bright and dark states was characterized in~\cite{Manovitz2017MSc}: for 1~ms exposure times and two-ion inter-ion distances exceeding 1.9~$\mu$m, the per-ion detection error rate is below $10^{-4}$, making detection errors a negligible contribution to infidelity.

The dominant SPAM contribution is imperfect state preparation. Combined 422~nm and 674~nm optical pumping prepares each ion in a single Zeeman sublevel of $S_{1/2}$. Rabi time scans on the 674~nm $S_{1/2} \leftrightarrow D_{5/2}$ transition, recorded on the same day as the gate data, reach a fitted flop amplitude of $0.997 \pm 0.003$ per ion (10 scans, 200 shots each). This amplitude upper-bounds the combined preparation and shelving-pulse error at $\sim$0.3\% per ion, or $\sim$0.6\% for two ions.

\subsection{Qubit dephasing}
\label{sec:supp-dephasing}

Ramsey interferometry on the 674~nm $S_{1/2} \leftrightarrow D_{5/2}$ quadrupole transition yields a coherence time $T_2 \approx 50 \pm 5$~ms in the same apparatus, limited by slow magnetic-field fluctuations and laser phase noise~\cite{Manovitz2022}. Over the gate durations used in this work ($T_g \approx 150$--$310~\mu$s), the resulting parity-contrast loss is $1 - e^{-T_g/T_2} \approx 0.3$--$0.6\%$, a minor but non-negligible contribution to the observed offset.

\subsection{Motional heating}

Sideband thermometry of the stretch mode yields a heating rate below $0.01$~quanta/ms, consistent with zero (Sec.~\ref{sec:experimental-setup} of the main text). Over the gate durations used ($T_g \approx 150$--$310~\mu$s), this corresponds to $\Delta\bar{n} < 0.003$ additional phonons, a negligible contribution to infidelity through the thermal factor in Eq.~\eqref{eq:fidelity}.

\subsection{Uncorrected amplitude-to-phase conversion}
\label{sec:supp-ampm}

The AOM phase response introduces an amplitude-dependent optical phase shift of up to $\sim$0.2~rad across the operating range (\figref{fig:aom_response}{b}). This amplitude-to-phase conversion is included in the forward simulations (Sec.~\ref{sec:supp-simulation}), and the close agreement between simulated and measured spectral curves in Figs.~\ref{fig:cardioid_spectrum_comparison} and~\ref{fig:supp_r23} confirms that the simulation accurately captures its effect on the IM3 tone powers. Nevertheless, the present DPD implementation corrects only the amplitude response. The residual phase distortion modifies the optical waveform on a sample-by-sample basis, potentially affecting both the amplitudes and phases of all spectral components. The full impact of this uncorrected amplitude-to-phase conversion on gate fidelity has not been independently characterized and may contribute to the observed fidelity offset.

\subsection{Spectral model limitations}

The photodiode-based fidelity estimate $\mathcal{F}_{\mathrm{est}}$ retains only the four dominant spectral components ($n = 0, 1, 2, 3$) and infers the IM tone phases from a cubic nonlinearity model (Sec.~\ref{sec:supp-fidelity-estimation}) rather than extracting them directly from the heterodyne measurement. Higher-order products (e.g., IM5 at $n = -1, 4$) are neglected; given that the IM3 tones are already $\gtrsim$20~dB below the gate tones, fifth-order products are suppressed by an additional $\sim$20~dB and contribute negligibly. The effect of relaxing the remaining assumptions has not been characterized and may account for part of the residual fidelity offset. In particular, AOMs are known to exhibit memory effects~\cite{Neumuller2024}, but their magnitude at the modulation bandwidths used here ($\sim$4~MHz) has not been quantified.

\subsection{Summary}

Table~\ref{tab:error-budget} lists the error sources characterized above. The quantified contributions account for $\lesssim$0.9--1.2\% of infidelity, leaving a gap of $\sim$0.6--0.9\% relative to the measured offset $|\delta| \approx 1.8\%$ that is not accounted for by the sources we have characterized.

\begin{table}[h]
\centering
\caption{Characterized contributions to the fidelity offset $\delta$.}
\label{tab:error-budget}
\begin{tabular}{lc}
\hline\hline
Error source & Estimated infidelity \\
\hline
State preparation (optical pumping) & $\lesssim$0.6\% \\
Qubit dephasing ($T_2 \approx 50$~ms) & 0.3--0.6\% \\
Motional heating ($\dot{\bar{n}} < 0.01$~q/ms) & $<$0.01\% \\
State detection (camera) & $<$0.02\% \\
\hline
Total (characterized) & $\lesssim$0.9--1.2\% \\
Measured offset $|\delta|$ & $1.8 \pm 0.3$\% \\
\hline\hline
\end{tabular}
\end{table}

Crucially, $\delta$ is a single constant shared across all drive amplitudes and DPD settings; it therefore does not affect the relative comparison between predistorted and uncorrected gates. The DPD-induced fidelity improvement is captured by the amplitude-dependent shape of the photodiode-derived fidelity curves, which track the measured data after the constant shift $\delta$. This confirms that the dominant amplitude-dependent error source is the IM3 distortion that DPD is designed to suppress.

\end{document}